**Neglecting Uncertainties Biases House-Elevation Decisions to Manage Riverine Flood Risks**

Mahkameh Zarekarizi[1*#], Vivek Srikrishnan[1], and Klaus Keller[1,2]

1 Earth and Environmental Systems Institute, the Pennsylvania State University, University Park, PA, USA

2 Department of Geosciences, the Pennsylvania State University, University Park, PA, USA

\* Corresponding author

Corresponding author email: mahkameh.zare@gmail.com

\# Now at Jupiter Intelligence

**Abstract**

Homeowners around the world elevate houses to manage flood risks. Deciding how high to elevate a house poses a nontrivial decision problem. The U.S. Federal Emergency Management Agency (FEMA) recommends elevating existing houses to the Base Flood Elevation (the elevation of the 100-yr flood) plus a freeboard. This recommendation neglects many uncertainties. Here we analyze a case-study of riverine flood risk management using a multi-objective robust decision-making framework in the face of deep uncertainties. While the quantitative results are location-specific, the approach and overall insights are generalizable. We find strong interactions between the economic, engineering, and Earth science uncertainties, illustrating the need for expanding on previous integrated analyses to further understand the nature and strength of these connections. Considering deep uncertainties surrounding flood hazards, the discount rate, the house lifetime, and the fragility can increase the economically optimal house elevation to values well above FEMA's recommendation.

# 1. Introduction

Floods affect billions of people worldwide[1]. 40 percent of natural disasters in the U.S. between 1900 to 2015 were floods[2]. Between 1970 and 2019, over 68 billion U.S. dollars have been claimed by National Flood Insurance Program (NFIP) policyholders[3]. The average damage to households has been over 30,000 U.S. dollars per event[3]. More than 100 million people live in the 100-year flood zone world-wide[4]. In the U.S., the population is disproportionately living in higher-risk areas with 41 million people living currently in a 100-yr floodplain[5]. This number is projected to increase to roughly 75 million people by 2100[5].

Flood risks can be reduced at the building level[6]. These approaches generally fall into three categories of wet-floodproofing, dry-floodproofing, and structure modification[7]. Dry-proofing prevents water from entering the building by closing the openings such as windows and doors or filling the basement. Wet-proofing allows water to flow inside the building, but reduces the vulnerability of the structure, for example, by moving valuable contents to higher floors[7,8]. Structural measures such as relocating, elevating, or demolishing a house in a flood zone are generally more effective for extreme floods[7,8].

Elevating a house can considerably reduce flood losses[7,9–11]. The U.S. Federal Emergency Management Agency (FEMA) recommends elevating houses in 100-year flood zones to at least the Base Flood Elevation (BFE) (the flood level with an annual exceedance probability of 1%) plus at least one foot of freeboard[12]. This recommendation provides, however, just a lower bound. Elevating a house above the FEMA's minimum requirement can be cost-effective[13]. How high to elevate a house is, however, a nontrivial decision problem. This motivates the following questions: (i) when does elevating a house result in monetary benefits exceeding the costs; and (ii) what height results in the highest net benefits? These questions are typically analyzed using Cost-Benefit Analyses (CBA). A CBA compares the investment cost (i.e. cost of elevating the structure) with the current (i.e. discounted) value of the expected benefits (i.e., the expected savings in future flood damages)[7,10].

The estimated costs and benefits are uncertain because they depend on uncertain inputs such as projected flood hazards, building vulnerabilities, discount rates, and the building lifespan[14–20]. For example, flood projection uncertainty arises from the uncertainties surrounding the choice of model structures, model parameters, model inputs, and realization of unresolved processes[21]. The house lifetime is uncertain because it is impacted by uncertain factors such as its structural durability, social acceptability, change in land value, and change in occupant needs[19,22]. Discount rates reflect the opportunity cost of spending money today rather than adding to investments[23]. A common approach to quantifying discount rates is to describe the observed opportunity costs and to analyze a relatively safe investment opportunity on decadal time scales such as U.S. Treasury bonds. The projected yields on these bonds are stochastic[20], resulting in dynamic uncertainty in potential investment yields and hence uncertain descriptive discount rates.

One technique for handling these uncertainties is to characterize various states by their probability of occurrence and assigning a Probability Density Function (PDF). This approach can be very useful to characterize aleatory uncertainties, but it can struggle to represent the effects of epistemic uncertainties[24,25]. Faced with sizable epistemic uncertainties, decision-makers do not always agree on a single PDF. This situation is referred to as "deep uncertainty"[26,27]. One approach to deal with deep uncertainty is to use alternative scenarios, for example by considering a set of plausible PDFs and to apply robust simulation methods that evaluate multiple competitive models or solutions and seek solutions that are capable of tolerating deviations from the conditions they were designed for[24,25,28].

Furthermore, conventional CBA typically focuses on a single objective: total discounted expected costs. However, stakeholders may have additional (and potentially conflicting) objectives. For example, homeowners may intrinsically value the reliability of avoiding flooding and the robustness of the strategy in the presence of deep uncertainty. Thus, analyzing the house elevation decision as a multi-objective problem can provide useful insights[29].

Here we use a multi-objective robust decision-making method[29] to analyze the house elevation decision problem. We identify important sources of (often deep) uncertainties, analyze their interactions, and characterize trade-offs between objectives. Previous work on the house elevation decision problem has provided valuable insights, but has been largely silent on the effects of uncertainties in the objectives and their potential trade-offs[10]. In general, we expand on the previous[8,30] work addressing the impacts of uncertainties on flood risk management decisions by refining the analyses of (i) deep uncertainties, (ii) the interactions on decision objectives, and (iii) the trade-offs between multiple stakeholder objectives. We show that the FEMA-recommended heightening policy is typically not optimal in a cost-benefit sense and often fails a cost-benefit test. We show that representing deep uncertainties can considerably change the projected risks and the choice of risk management strategy. We provide generalizable insights about the effects of the choice of monetary discount rate on decision-making for longer-term projects and the possible tensions between adopting a descriptive vs. prescriptive approach in choosing a discount rate. Our analysis demonstrates how analyzing these decisions requires a tightly integrated and transdisciplinary approach, as the decision is driven by complex interactions between uncertainties surrounding the Earth-, social-, and engineering systems.

## 2. Results

2.1. Impacts of uncertainties on objectives

We demonstrate the approach for a case study in Selinsgrove, a rural location in Pennsylvania (PA) in the Eastern U.S. (Supplementary Note 1). We focus on four strategies: (i) repairing flood damages as they occur, (ii) elevating the house to FEMA's minimum recommended height, (iii) elevating the house to the cost-optimal heightening strategy neglecting-uncertainty, and (iv) elevating the house to the optimal height considering uncertainty. We consider four sources of uncertainty (Supplementary Figure 1). First, we quantify the chance of being flooded in any given year. Ignoring the considered uncertainties can drastically underestimate flooding probability by a factor of 5% (Figure 1d-e). The downwards bias is exacerbated for floods with higher return periods. This underestimation drives also an underestimation of Annual Expected Damages (EAD) (Figure 1a-c).

Second, we quantify the uncertainty surrounding projected discount rates using past observations of discount rates. Results show that neglecting the uncertainty surrounding future discount rates can drastically underestimate future damages (Figure 2). Uncertainty in future discount rates increases the Net Present Value (NPV) of projected flood damages (Figure 2a). The discount rate is an important factor in this assessment, as it translates futures costs to today[20]. Flood risk management studies often use a prescriptive approach with a constant and perfectly known future rate, for example, 4% per year[10]. In contrast, other studies adopt a descriptive approach and adopt multiple descriptive discount rates[31,32].

In the prescriptive approach, the discount rate is a choice and can be treated as certain. In the descriptive approach, the discount rate depends on the interest rate available in actual investment markets[33]. In other words, descriptive discount rates stem from the time value of money, which is related to the interest an alternative investment would have yielded. For the house elevation problem, relevant and relatively safe investment opportunities on decadal time scales are U.S. Treasury bonds. The yield on these bonds are stochastic. Thus, this type of discount rate is uncertain in that it captures stochasticity in the underlying net bond rates. Hence, quantifying the expected discount rate over the lifetime of the investment is important for understanding the expected yield on an alternative investment in treasuries, which affects the ability of the decision-maker to get richer over time through the alternative investment than by investing in elevation today. Therefore, since the discount rate depends on uncertain projections, it is uncertain[20]. Adopting a fixed and perfectly known discount rate can provide useful insights, but is silent on the effects of uncertainty about future discount rates in the framework of a descriptive model that is consistent with observations and does account for key effects of projection uncertainty. Whether one chooses a prescriptive or descriptive approach to identify a discount rate depends on a range of methodological, economic, political, legal, and philosophical questions (see, for example, the discussions in[34,35]). One key problem with adopting a prescriptive approach is that it can lead to inconsistent choices[34]. As an example, the FEMA recommendation of 7 % per year[36] is inconsistent with the observed and projected alternative investment opportunities a homeowner in the United States currently has access to (Figure 2). A homeowner may choose to adopt the FEMA recommendation of 7% per year to analyze the decision to elevate a house but faces a rather different discount rate when making decisions about alternative investments, for example, whether to buy government bonds.

Third, we quantify the uncertainty surrounding the flood vulnerability of the building[37,38]. Common vulnerability models are depth-damage functions that quantify the damages for a certain depth of water in a house. These damage models are deeply uncertain in the sense of model structure, as demonstrated by the divergence of the model predictions[38].

Finally, we sample the uncertainty surrounding the house lifetime. The house lifetime is uncertain because it is impacted by uncertain factors such as its structural durability, social acceptability, change in land value, and change in occupant needs[19,22]. Flood risk studies often use a deterministic value between 30 to 100 for residential buildings' lifetime and ignore the surrounding uncertainty[7,10,31,39].

We quantify the effects of these four uncertainties. For flooding probability and house lifetime we use a PDF to represent the uncertainty. For the discount rate and damage model, we consider them deeply uncertain and use multiple model structures and PDFs to quantify their uncertainty.

We start with analyzing a hypothetical 1,500 ft$^2$ house with a worth of $300K and with the lowest floor at four feet below the BFE. Total costs include investment cost plus the net present value of expected damages. If this house is not elevated, total costs could be more than the house value ($V$). With 90% probability, these costs are between $0.17V$ and $1.61V$ with an expected value of $0.68V$ (Figure 3a). Total costs are $0.67V$ if the house is elevated by 14 feet (ten feet above the BFE). The optimal elevation that minimizes the expected total costs is 8.8 feet (4.8 ft above the BFE). At this heightening strategy, expected total costs are $0.59V$. These costs are less than the house value with high probability.

Ignoring uncertainty changes the optimal elevation with respect to the CBA (Figure 3a). Ignoring uncertainty, the total cost without elevating is nearly $0.68V$. Ignoring uncertainty underestimates the expected damages and the resulting cost-benefit analysis suggests not to elevate the house. Considering uncertainty changes the decision to elevate the house by 8.8 ft. Considering uncertainties leads to a higher optimal elevation because it increases the expected damages while leaving the costs unchanged. By adopting the recommendation that neglects uncertainty, the house owner risks $203K (NPV), which is considerably higher than the cost of elevating the house (i.e. ~ $152K). The FEMA recommendation suggests elevating this house by at least 5.5 feet (the minimum freeboard recommended by FEMA in Selinsgrove is 1.5 feet). This costs the homeowner $145K. Implementing FEMA's recommendation reduces the expected total costs from $0.68V$ to $0.65V$. However, this strategy is suboptimal with respect to the benefit-to-cost ratio (BCR) (Supplementary Figure 2).

In summary, implementing the strategies derived when neglecting uncertainty, following FEMA, and considering uncertainty costs the homeowner zero, $0.48V$, and $0.5V$, respectively. The NPVs of the expected total costs of these strategies are $0.68V$, $0.65V$, and $0.59V$, respectively. Thus, implementing the strategy recommended by the considering-uncertainty assumption costs marginally more but these extra costs are more than offset in future damages.

Next, we evaluate the BCR to ensure that the implemented strategy passes the cost-benefit (CB) test. If the homeowner elevates the house by more than five feet, the benefits are expected to exceed the costs (strategy passes the cost-benefit test) (Supplementary Figure 2). The expected BCR of the optimal strategy is 1.16. The optimal strategy is expected to pass the CB test. Ignoring uncertainty implies that elevating this house is never cost-effective. The FEMA-recommended strategy has a BCR of 1.04 and passes the CB test (Supplementary Figure 2).

Another homeowner's objective may be to maximize reliability, the probability of no flooding over the house lifetime. Expected reliability is more than 50% for all heightening strategies greater than four feet (Supplementary Figure 3). If the house is not elevated, the reliability is 16%. In other words, there is an 84% chance that it will be flooded at least once over its lifetime. This chance of flooding drops to 22% if the house is elevated to the optimal elevation under uncertainty. The expected reliability of the FEMA-recommended strategy is 60%. Ignoring uncertainties overestimates the reliability and underestimates the chance of being flooded. This leads to a false sense of security.

A robust decision performs sufficiently well (depending on the robustness criterion) across many plausible alternative future conditions, at the potential cost of worse performance in the expected future[28]. We quantify robustness using a satisficing metric[29,40]. Specifically we evaluate the robustness as the fraction of parameter samples (each referred to as a state-of-the-world or SOW) for which one or all objectives are within the decision-makers' acceptable ranges (i.e. greater than one for the BCR, [0,0.75] for the ratio of the total cost to house value, and [0.5,1] for reliability). If the house is elevated to five feet or more, 40% of SOWs lead to an acceptable BCR (Figure 3b). If the homeowner decides not to elevate the house, none of the SOWs are within the acceptable range of reliability and only 65% of SOWs are within the acceptable range of total cost. However, if elevated by 10 feet or more, the robustness of reliability grows to 90%. Overall, the decision not to heighten the house satisfies all criteria in 0% of scenarios, the FEMA-recommended strategy in around 14% of scenarios, and the economically optimal strategy in 37% of scenarios (Figure 3b).

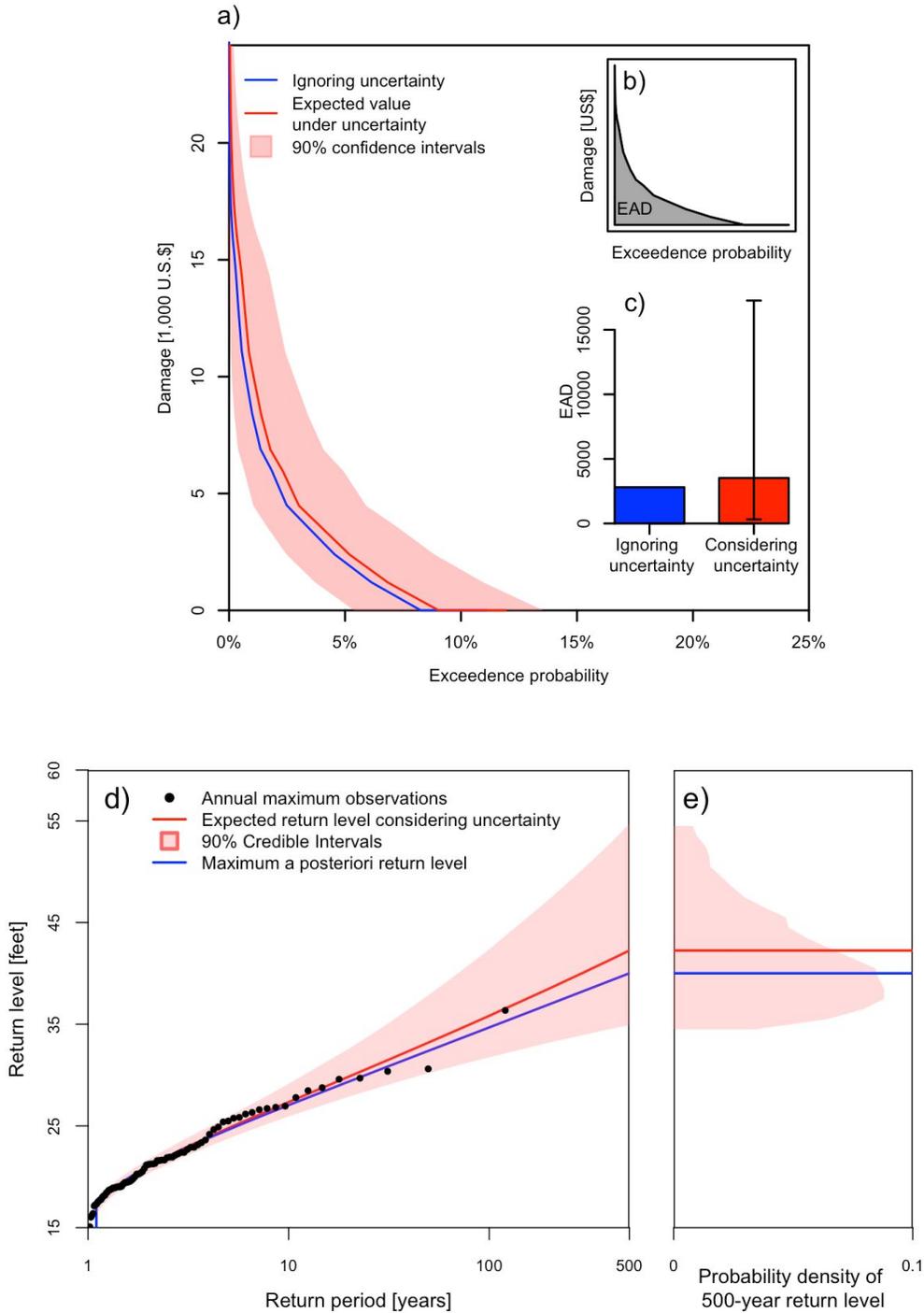

Figure 1: Characterization of flood hazards and damages. Expected Annual Damages (EAD) is the area under the Exceedance Probability Loss (EPL) curve that indicates damage versus flood probability (b). EPL curves under the considering-uncertainty (red line and bounds) and ignoring-uncertainty (blue line) assumptions are compared in panel a. The resulting EADs are compared in panel c. The shaded red area (in a) indicates the 90% credible intervals of the considering-uncertainty assumption. The narrow line on the red bar indicates the range of uncertainty in EAD. Return levels of the two assumptions are compared in panel d. Panel e exhibits the comparison for 500-yr flood

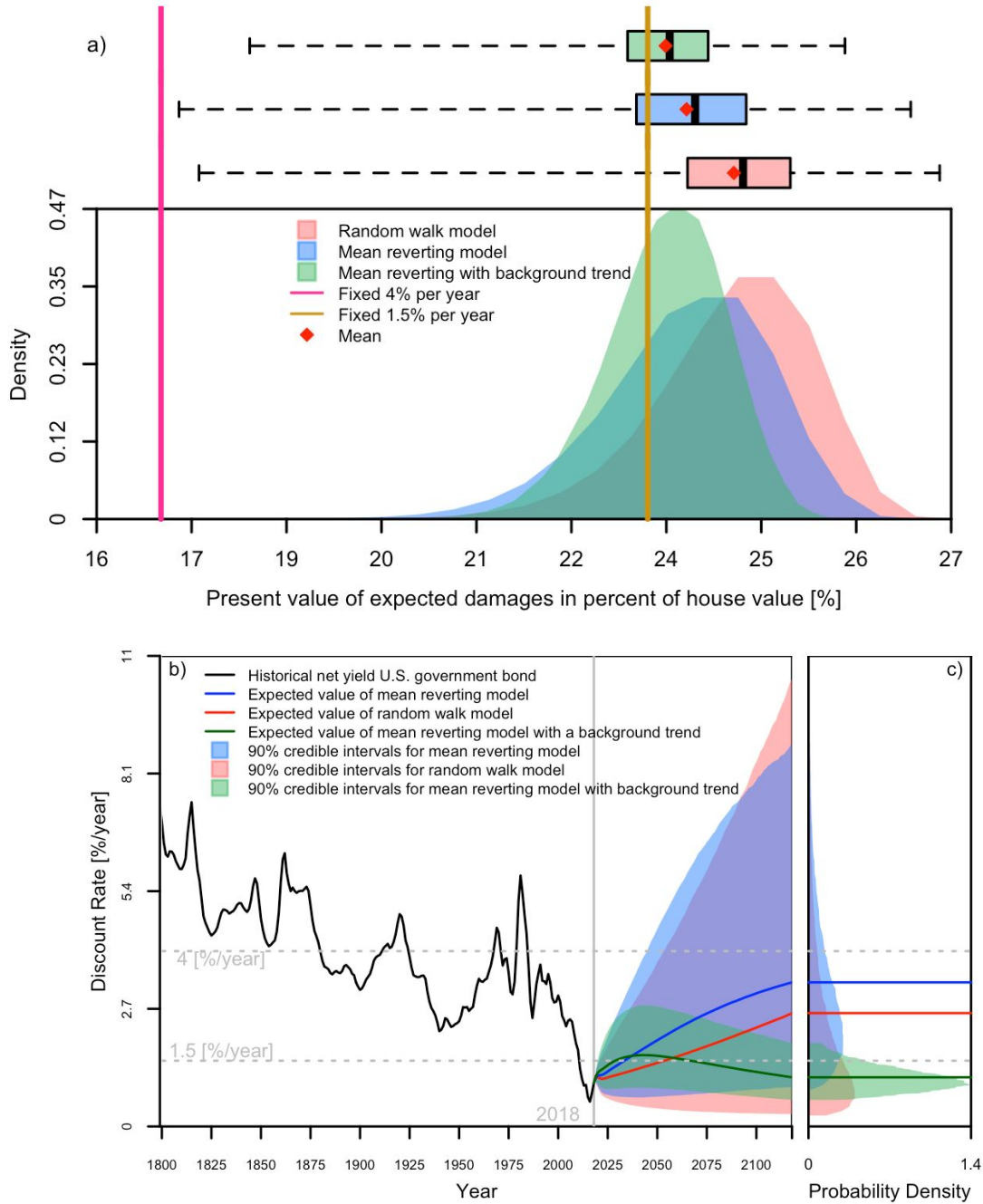

Figure 2: Impact of different discount rate models on estimates of the net present value of expected damages for the hypothetical house (1,500 ft$^2$ with a worth of $300K and with the lowest floor at four feet below the Base Flood Elevation). Box plots show the dispersion of the damage estimates for the three considered stochastic models (b) Historical (1800-2018) and projected discount rate time series. The shaded areas indicate the 90% credible interval of projected discount rates. The whiskers extend to the data extremes. Boxplot centerline is the median

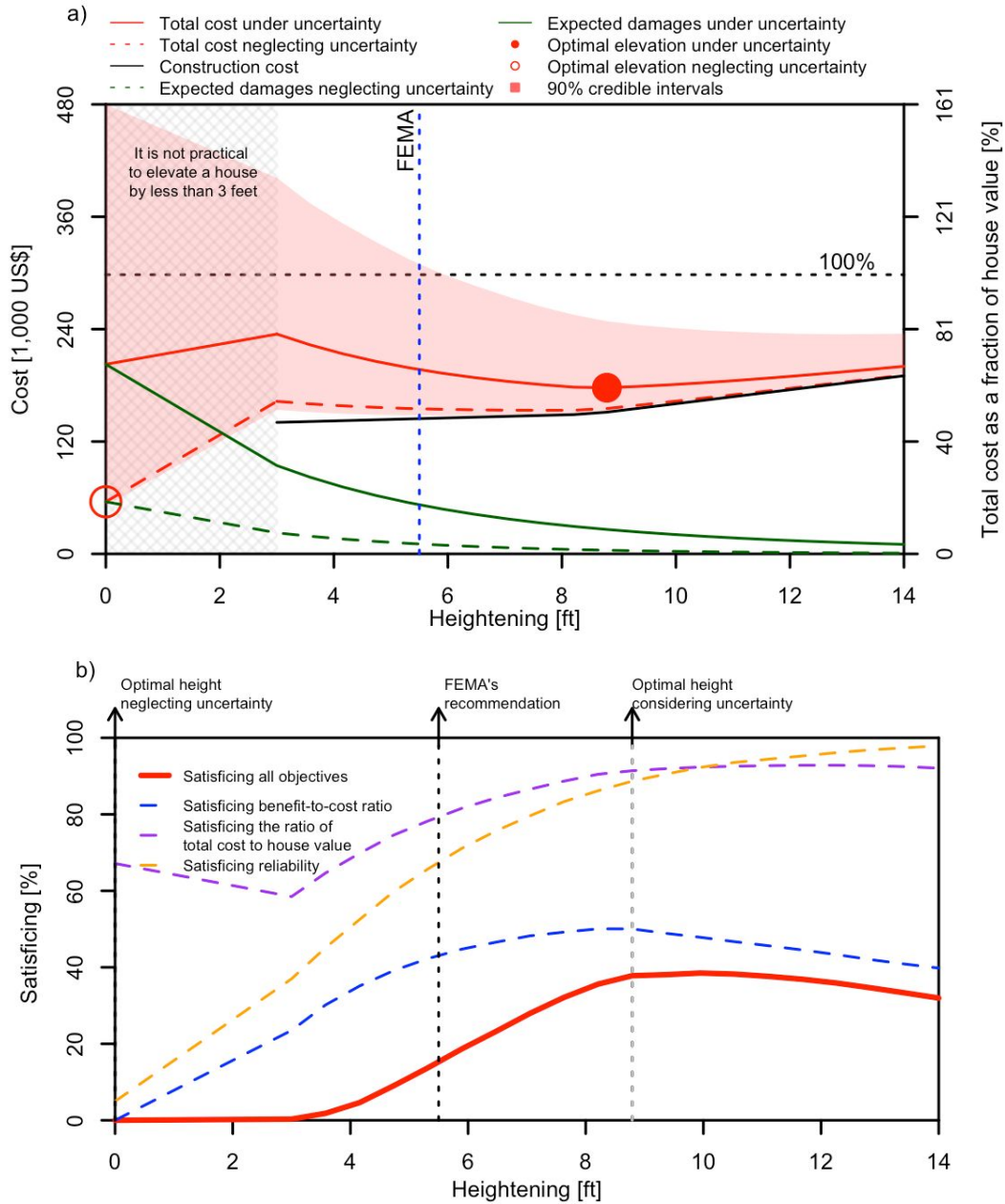

Figure 3: Cost-benefit and robustness analysis of heightening strategies. (a) Total cost and the optimal elevation under assumptions of ignoring-uncertainty (dashed magenta line and the hollow point) and considering-uncertainty (solid magenta line, shaded bounds, and the filled point). 1,500 ft$^2$ house with a worth of $300K and with the lowest floor at four feet below the Base Flood Elevation. Under the ignoring-uncertainty assumption, the house lifetime and discount rate are assumed to be 30 years and 4% per year, respectively. The vertical line indicates the FEMA-recommended heightening strategy. The hatched gray area on the left refers to elevating the house by less than three feet which we assume is impractical in this study. (b) Robustness of heightening policies. Robustness of different objectives are shown by dashed lines. The solid red line indicates the robustness of all objectives

## 2.2. Trade-off Analysis

The considered objectives show strong trade-offs. Reliability and upfront costs are two competing objectives in the house elevation decision (Figure 4 and Supplementary Figure 4). It is infeasible for the considered case to achieve perfect reliability with zero upfront costs (star in Figure 4). A small heightening strategy has a low upfront cost and low reliability. A large heightening corresponds to relatively high reliability but requires high investments that might not be affordable. Ignoring uncertainty moves the estimated Pareto front into the infeasible zone in the case when the uncertainties are considered. One key driver for this effect is that considering uncertainty reduces reliability (Supplementary Figure 3).

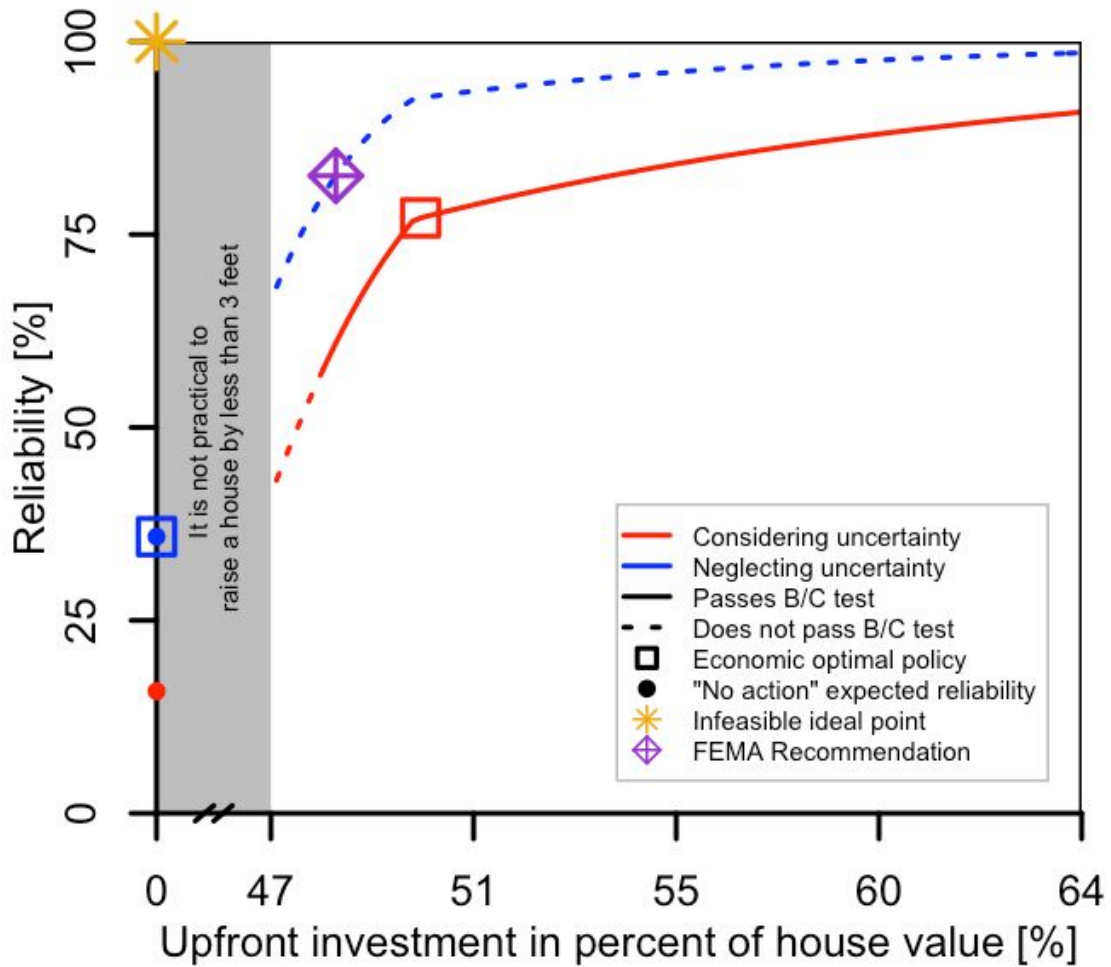

Figure 4: Trade-offs between the upfront cost and reliability with and without considering uncertainties. The trade-off considering-uncertainty and ignoring-uncertainty are shown in red and blue, respectively. Along each line, the dashed parts indicate that the policy does not pass the cost-benefit test (i.e. the benefit-to-cost ratio is less than one). Heightening policies of 0-3 feet are blocked by the gray area as we assume that it is impractical to elevate a house by less than three feet. The "not elevating" policies are shown by dots and the optimal elevations are shown by squares

## 2.3. Uncertainties that drive the variance of projected damages

What are the most important uncertainties and how do these uncertainties interact? We analyze these questions using a global sensitivity analysis[41]. This approach quantifies the relative importance of uncertainties from individual inputs or parameters (first-order sensitivities) or from their interactions (i.e. second-order sensitivities, if the variance in the output results from interactions between two inputs). We analyze, as an example, the drivers of uncertainty surrounding projected damages. There are two sources of deep uncertainty including the damage model with two options and the discount rate model with three options. Thus, there are a total of six scenarios.

For all scenarios, the expected damages are sensitive to a complex interplay of uncertainties surrounding the discount rate, damage function, house lifetime, and flood frequency (Supplementary Figure 5). The shape parameter for the flood distribution has the largest effect on the damage uncertainty. This is, perhaps, expected, as the expected probability of flooding in any given year has a direct impact on the expected annual damages and consequently on the lifetime expected damages. After the flood frequency model parameters, lifetime and damage model uncertainties play the most important roles. The dominant second-order interactions are between the frequency model parameters. For the most likely scenario (Supplementary Table 1), out of five statistically significant (at 95% confidence level) second-order interactions, two are with the house lifetime uncertainty (Figure 5). Furthermore, for the majority of scenarios, there is a statistically significant second-order interaction between the discount rate and lifetime uncertainty (Supplementary Figure 5). When houses have longer lifetimes, different discount rate models diverge even more (Supplementary Figure 6). For such houses, the discount rate model structure plays an even more important role. For houses with a lower lifetime, the discount rate models do not result in considerably different projections.

Sensitivity analysis also allows us to assess the relative importance of different model structures in factors that are deeply uncertain. Thus, we assess the relative importance of the discount rate model structure and the depth-damage function structure (Supplementary Figure 7). By considering deep uncertainties, the depth-damage model structure becomes more statistically significant and the frequency model parameters become less significant.

These results are based on a sample house that is worth $300K, has 1,500 ft$^2$ and is four feet below the BFE. We analyze the impacts of different exposures and re-evaluated these sensitivities for a set of hypothetical houses (Supplementary Figure 8) under the most likely scenario. For all the cases, the uncertainties in the flood probability, house lifetime, discount rate, and depth-damage function are statistically significant drivers of the variance in projected damages, regardless of house exposure factors. Flood frequency model becomes less important and discount rate uncertainty becomes more important for houses that are farther below the BFE.

One important takeaway is that neglecting discount rate uncertainty can considerably underestimate the damages. If a fixed discount rate is used, its value becomes the most important factor that explains the variance in the damages (Supplementary Figure 9). However, if an uncertain stochastic model is used, its uncertainty becomes less important (Figure 5) and the model choice has much less of an effect on the projected damages (Supplementary Figure 7). This is largely because the stochastic discount models do not produce very different projections over the house lifetime.

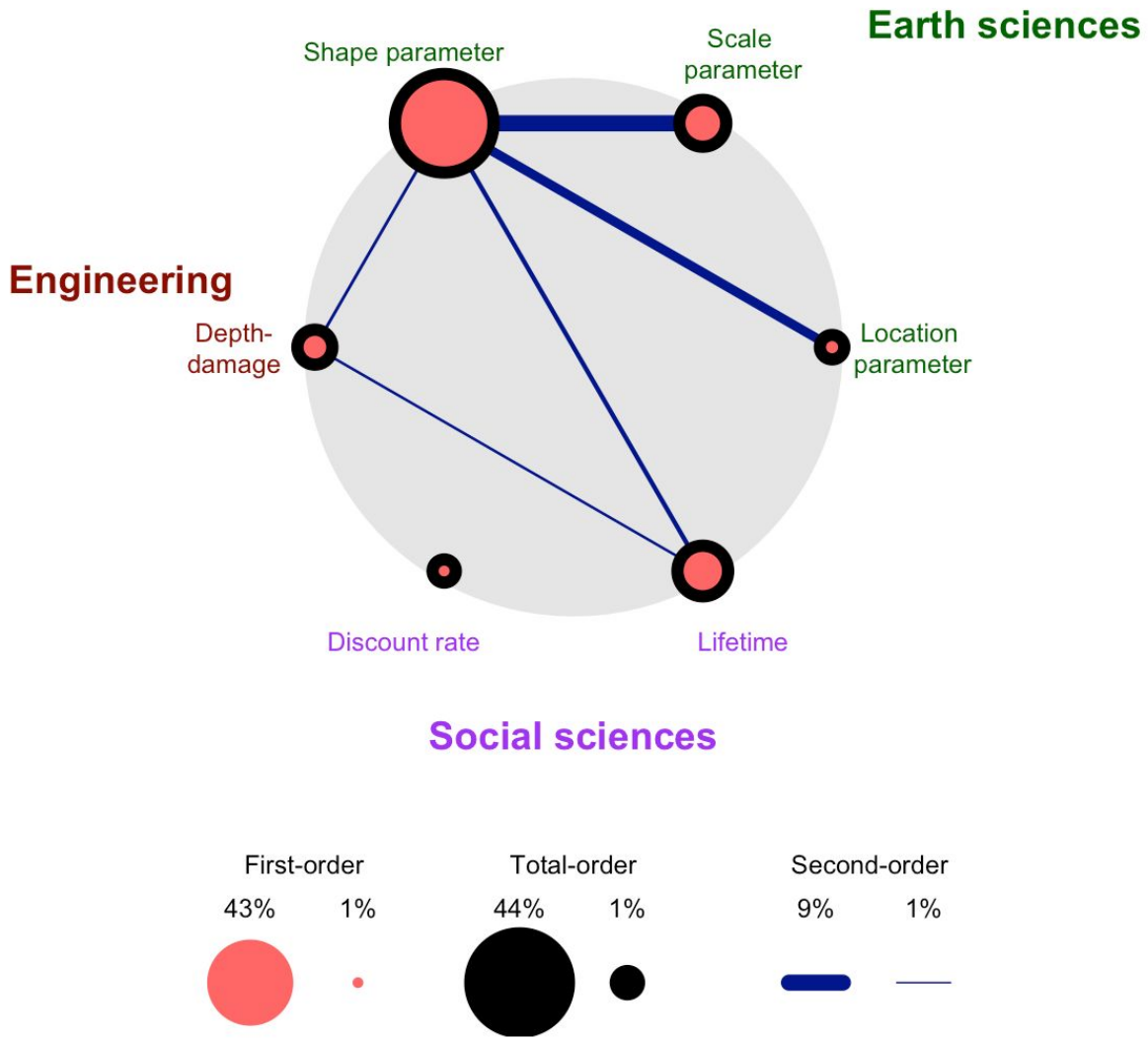

Figure 5: Sensitivity of lifetime expected damages for a typical (1,500 ft$^2$ with worth of $300K) house in Selinsgrove, PA to the considered uncertainty sources. Results are from a global sensitivity analysis. Salmon circles are proportional to the individual sensitivity of each source. Lines' thicknesses are proportional to the relative importance of the interactions between the two sources. The black circle diameter is an indicator of both. Indices that are not significant at a 95% confidence level are not drawn

## 2.4. Effects of house exposures characteristics

Houses vary in terms of exposure, as measured by factors such as house size, value, and the lowest floor elevation. The analysis, thus far, focused on objectives and uncertainties for a single sample house. In this section, we address the effects of house exposure factors on the mitigation decision. To this end, we analyze the multi-objective robust decision framework described above for 1,000 hypothetical houses (Supplementary Table 2) that sample exposure factors. Ignoring uncertainty decreases the optimal elevation for all considered houses (Supplementary Figure 10). For 68% of the houses, the optimal elevation is higher than FEMA's recommendation (Supplementary Figure 11). On average, the optimal elevation is approximately one foot higher than FEMA's recommendation. This means that if the hypothetical house owners raise their houses by a few feet higher than the FEMA-recommended elevation, they save more in future damages. For around 23% of the buildings, the optimal elevation is zero, but FEMA recommends elevating them. In all of those houses, FEMA's recommendation would not pass the cost-benefit test. In about 8% of the houses, the optimal elevation is less than FEMA's recommendation. In almost all of them, FEMA's recommendation does not pass the cost-benefit test. In all the houses with different elevations, sizes, and values, the optimal elevation passes the cost-benefit test. However, in only 38% of houses, the FEMA-recommended strategy passes the cost-benefit test.

## 3. Discussion

A considerable fraction of the global population lives in floodplains. Homeowners in these floodplains are making nontrivial decisions about how to manage flood risks. One common flood risk mitigation strategy is to elevate existing buildings in flood-prone regions. FEMA recommendation suggests elevating at-risk houses to at least one foot above the Base Flood Elevation (BFE), the water elevation associated with the 100-year flood[12]. This recommendation still leaves open the question of whether (and if so, by how much) to elevate the houses.

This house elevation problem is typically addressed in a single objective cost-benefit framework[10]. Traditional approaches seek an optimal strategy that minimizes the total cost, which is the net present value of expected damages plus the investment cost. Stakeholders can, however, have multiple objectives such as maximizing the benefit-cost ratio, minimizing the upfront cost, or maximizing robustness. Stakeholders can differ in their relative preferences regarding these objectives and their constraints. For example, some stakeholders may choose to increase robustness by investing in a higher house elevation while others may choose not to. We quantify and assess these objectives and their trade-offs.

Analyses of the house elevation problem often neglect key uncertainties. Estimating the total cost requires projections of the flooding probability, the damage function, the monetary discount rate, and the expected house lifetime. Traditional approaches often adopt deterministic values for these inputs. For example, many studies choose one or two fixed discount rates[7,10,31]. This neglects key aspects of the uncertainty surrounding potential investment returns, that are available to homeowners as an alternative to elevation. For the damage model, a typical choice is the depth-damage functions by FEMA[42]. For flooding probability, the standard approach is to use a probability distribution with perfectly-known distribution parameters[30]. Ignoring these uncertainties can bias the projected expected damages. This, in turn, can lead to drastic changes in the projected trade-offs and acceptable decisions. We quantify the considerable uncertainty surrounding the projected flood frequency parameters, house lifetime, projected discount rate, and the depth-damage function and show how these uncertainties impact the discounted expected damages. We demonstrate how the FEMA recommendation for heightening often fails a costs-benefit test and can typically be improved on. Currently, FEMA's recommendation is only based on the flood zone and elevation with respect to the BFE. Our findings suggest that taking house characteristics such as house value, house size, and initial elevation into account can improve outcomes. The owners of the hypothetical homes in our study can save in future damages if they raise their houses a few feet above FEMA's recommendations.

Our study is subject to several caveats that point to future research needs. First, this study focuses on a single decision lever of elevating a house (Supplementary Figure 1) and does not consider additional decision levers such as purchasing flood insurance or buy-outs[43-45]. Considering the effects of such additional levers poses highly relevant questions, but is beyond the scope of this analysis.

A second caveat arises from the still limited treatment of uncertainties. For example, our study neglects uncertainties surrounding future climate change (the adopted flood hazard model is stationary), the elevation costs, the lowest floor elevation of the house (depending on the Digital Elevation Model or survey data), structure type and material, nature of the watershed, and distance of the structure from the stream. Third, we adopt elevation cost estimates from a study applied to Louisiana[46]. An interesting expansion of our study would be to consider location-based cost estimates and the surrounding uncertainties. Fourth, we consider a one-shot decision about elevating an existing house and neglect the option to postpone the elevation and neglect changes in house value after elevation. The analysis is hence conditional on a previous decision: whether to build a house higher than FEMA's recommendation in the first place[13,31,42]. Designing new buildings with elevations above FEMA's recommendation can be cost-effective[13].

Flood risk mitigation recommendations and strategies vary across countries depending on various factors including governmental strategies and homeowners' flood risk perceptions[47–49]. The framework presented in this paper can be applied to cases outside the U.S. with appropriate changes, for uncertainty quantifications that are location-dependent. For example, while the depth-damage function depends on the location and building type, our overall approach to uncertainty quantification and trade-off analysis can still be applied (Supplementary Note 2). Last but not least, the mechanisms driving the flood hazard vary across locations. We consider just fluvial flooding in a stationary setting while other locations are exposed to different and nonstationary flood types (e.g., coastal storm surges[18,50]). These cases require a much more sophisticated characterization of projected flood hazards (see, for example refs[51,52])

In summary, we identify the key drivers of poor outcomes in the decision of elevating a house to manage flood risks. What seems like a simple risk mitigation decision can turn rather complex, once deep uncertainties and their interactions are considered. Our findings suggest that accounting for uncertainties in the discount rate, the depth-damage functions, and house lifetime can be fruitful avenues to improve this decision.

## 4. Methods

### 4.1. The framework

We use a Multi-Objective Robust Decision Making (MORDM) framework to analyze the house elevation decision[29] (Supplementary Figure 1). Exogenous uncertain factors in our framework are flooding frequency, discount rate, depth-damage curve, and house lifespan. The decision lever (i.e. actions that the decision-maker can take) is heightening (i.e. the added height to the house). We consider five objectives: (1) minimizing the total costs, (2) maximizing the benefit-to-cost ratio, (3) minimizing the upfront cost with respect to the initial value of the house, (4) maximizing reliability (i.e. the probability of no floods during the house lifetime), and (5) maximizing the robustness of the design to deviations from the best-guess parameters[28].

The closest U.S. Geological Survey (USGS) gage to Selinsgrove is USGS gage 01554000 collecting water data at Susquehanna River at Sunbury, Pennsylvania. Daily discharge data at this location are available for the period of 1937 to 2019 but daily gage height data are limited to 2000-2019. Thus, in order to take advantage of the rather long record of discharge data, we use the USGS stage-discharge rating curve for this location to convert discharge to gage height.

**4.2. Uncertainties**

We use different robust methods to quantify the uncertainty of each factor, depending on the nature of that factor or previous research findings about the uncertainty of the variable. For example, in the cases of discount rate and the depth-damage function, we use multiple competitive models[24]. Below, we review our method of quantifying flooding probability, discount rate, damage curve, and the house lifetime, respectively.

We quantify the uncertainty surrounding flood probabilities using a Generalized Extreme Value (GEV) distribution combined with a Markov Chain Monte Carlo (MCMC) sampling for parameter estimation. Using the maximum *a posteriori* estimates of GEV parameters (as opposed to the full parameter sample) underestimates the flood hazard (Figure 1d-e). This effect is driven by the right-skewed nature of the return level distribution where the mode is smaller than the mean (Figure 1b). This underestimation drives also an underestimation of the Annual Expected Damages (EAD) (Figure 1a-c). EAD is the area under the Exceedance-Probability Loss (EPL) curve that represents the damages versus exceedance probability (Figure 1a). Comparing the EPL curves neglecting and considering uncertainty (Figure 1a) illustrates how ignoring uncertainty underestimates EAD.

The GEV distribution is used for modeling annual maximum daily water level (maximum daily water level in the course of a year) and is recommended by FEMA[53]. We hence approximate the annual maximum floods distribution using a GEV distribution (Supplementary Note 3). To estimate the GEV parameters, we use MCMC sampling within a Bayesian framework. We adopt the MCMC sample with the highest posterior probability samples as the "best guess" estimate of that parameter. To account for the uncertainty of flooding frequency, we consider the full ensemble of samples.

The Cumulative Distribution Function (CDF) of GEV (i.e. the probability of annual maximum water level; AMWL; not exceeding level $h$) is

$$Pr(H \leq h) = exp\{-[1 + \xi(\frac{h-\mu}{\sigma})]^{\frac{-1}{\xi}}\}, \tag{1}$$

where $H$ is a random variable representing AMWL. $\mu$, $\sigma$, and $\xi$ are location, scale, and shape parameters, respectively. Prior distributions for $\mu$, $\sigma$, and $\xi$ are normal distributions centered at zero. For posterior sampling, we use one MCMC chain initialized at five, one, and 0.1. Our sample size is 50,000 (Supplementary Note 4).

For discount rate, we expand on previous work[20] and quantify the uncertainty surrounding the projected rates using the observed record and time-series models. The observed historical discount rates are highly stochastic (Figure 2b). To account for deep model structural uncertainty, we follow previous work[20] and consider three autoregressive models, fitted to the logarithms of the discount rates, as there is no historical evidence of negative discount rates in the U.S. reflecting deep model structural uncertainty. Following ref. ([20]), the first model is a random walk and the second model is mean-reverting. We additionally consider a model with a background linear trend (on the log-scale). Accounting for this discount rate uncertainty results in a higher discount factor[20,54] $F_t$ and increases the net present value of projected benefits and costs (Figure 2a). This is an essential feature of stochastic discount rate models compared to using a single expected rate[20].

We estimate uncertain discount rate dynamics using an extension of the data from ref.[20]. As in that paper, we obtained estimates of expected inflation from a ten-year moving average of Livingston Survey Consumer Price Index (CPI) forecasts[55]. We subtract these estimates from annual nominal yields on 20-year Treasuries[56] to produce a series of historical discount rates. We follow ref. ($^{20}$) by then converting these rates to their continuously compounded equivalents and using a three-year moving average to smooth short-term fluctuations. The resulting discount rate time series, denoted $d_t$, is shown in Figure 2b.

Our discount rate models are autoregressive AR(3) time series models fit to this data, which maximizes the Akaike Information Criterion, AIC[57]. We use logarithms of the discount rates to ensure that the time series remains positive, due to the lack of evidence of negative rates in the U.S. Following ref.[20], we consider three models, reflecting deep model structural uncertainty. The first model is a random walk,

$$ln(d_t) = \rho_{t-1}d_{t-1} + \rho_{t-2}d_{t-2} + \rho_{t-3}d_{t-3} + \varepsilon, \quad \Sigma_t\rho_t = 1. \qquad (2)$$

The second model is mean-reverting with constant mean,

$$ln(d_t) = \eta + \rho_{t-1}(d_{t-1} - \eta) + \rho_{t-2}(d_{t-2} - \eta) + \rho_{t-3}(d_{t-3} - \eta) + \varepsilon, \; \Sigma_t\rho_t < 1. \qquad (3)$$

The third model is a mean-reverting model with trend,

$$ln(d_t) = \eta + \beta t + \rho_{t-1}(d_{t-1} - (\eta + \beta(t-1))) + \rho_{t-2}(d_{t-2} - (\eta + \beta(t-2))) + \qquad (4)$$
$$\rho_{t-3}(d_{t-3} - (\eta + \beta(t-3))) + \varepsilon, \quad \Sigma_t\rho_t < 1.$$

We show the estimated coefficients for all three models in Supplementary Table 3. The random walk and mean-reverting models have AIC values (Supplementary Table 4) which are statistically equivalent, as AIC differences less than 2 indicate similar levels of evidence for the compared models[58]. The background trend model has stronger support based on AIC[58], but a similar Bayesian Information Criterion (BIC) value to the mean-reverting model with constant mean[59]. As a result, we include all models in our analysis.

Depth-Damage functions translate flooding to its economic impacts[38]. They determine the susceptibility of entities at risk to floods and are key to damage estimation[17,60,61]. Depth-damage functions estimate potential damages for a certain amount of water (usually in the form of depth) in a house. There is a wide variety of published sources to obtain these curves[60]. Depth-damage functions are uncertain and we hence adopt a probabilistic treatment[38,60,62].

A common source of depth-damage functions in damage assessment studies in the U.S. is Hazard U.S. (HAZUS) provided by FEMA. In an attempt to aggregate various depth-damage curves, the Joint Research Centre (JRC) of the European Commission's science and knowledge service presented consistent global depth-damage functions[63]. They provide a depth-damage function for North America which aggregates various damage functions. All of these functions are derived from HAZUS.

To account for the depth-damage function uncertainty, studies often use multiple functions[16,46]. Other studies have used parametric distributions to quantify the damage model uncertainty[17]. A recent study addresses the validity of depth-damage function and provides further evidence on the uncertainty of these functions[38]. This study proposes that at a given depth, the damages follow a Beta distribution. Unfortunately, these probabilistic depth-damage functions are provided only up to eight feet, not enough for our study. Thus, we rely on previous studies and use two different depth-damage functions where each function has a uniform uncertainty bound around it[16,41]. We use two damage functions to represent the deep uncertainty in the damage curve. We represent the uncertainty of each function by assuming a uniform uncertainty of 30% around the curve[41]. Supplementary Figure 12 presents both curves and the uncertainty around each model.

It is crucial to estimate the anticipated lifetime of a structure for mitigation decisions[19]. The lifespan of a house is uncertain. The lifetime of a building is impacted by uncertain structural and social factors[19,22]. Many flood damage studies do not address the actual lifetime of a building and assume a typical value (i.e. 30 or 50 years)[7,10,19]. These studies ignore the surrounding uncertainty[7,10,31,39]. To the best of our knowledge, this is the first time that house lifetime uncertainty is considered in a flood mitigation study.

A study based on U.S. residential building stock data (provided by the U.S. Census Bureau under the 2009 American Housing Survey microdata) finds that the average residential building lifetime is 61 years with a standard deviation of 25 years[19] (Supplementary Figure 13). With 90% confidence, lifetime is expected to be between 21 and 105 years[19]. The distribution of building lifetime is best represented by Weibull distribution with shape and scale parameters of 2.8 and 73.5, respectively. In this study, we use the model suggested by that paper to quantify the uncertainty of house lifetime. We compare this distribution with previously published literature in Supplementary Figure 13. We adopt the Weibull distribution for the "considering uncertainty" assumption and the fixed value of 30 years for the "ignoring uncertainty" assumption.

### 4.3. Objectives

The first objective is the ratio of the upfront cost (cost of elevating the house) to house value ($O_{1h} = \frac{C_h}{V}$), where $V$ is the current value of the house (before elevating) and $C_h$ is the cost of elevating the building by $h$ feet. The cost of elevating a single-family house is interpolated from the Coastal Louisiana Risk Assessment Model (CLARA)[46]. According to this model, the unit cost of elevating a house by 3-7, 7-10, and 10-14 feet is $82.5, $86.25, and $103.75 per square feet with a $20,745 initial fee. The initial fee includes administration, survey, and permits. Supplementary Figure 14 depicts the interpolated construction costs for three hypothetical 1,000-, 2,000-, and 3,000-square feet houses.

Total cost ($O_{2h}$) is the upfront cost of lifting a house (by $h$ feet) plus the present value of lifetime expected damages (LED) if elevated by $h$ feet. LED is a function of expected annual damages (EAD) and is calculated by

$$LED_h = \sum_{t=0}^{n} EAD_h * F_t , \qquad (5)$$

where $EAD_h$ is the expected annual damages when a house is elevated by $h$ feet. $n$ is the house lifetime, and $F_t$ is the discount factor at year $t$.

Previous studies have either substituted EAD with NFIP insurance premiums[10] or calculated the expected damages[7,8]. The former method implies that NFIP premiums reflect the actual risk. However, NFIP was designed to subsidize the cost of flood insurance on existing houses[45,64,65] and is not risk-based especially for structures that were built before the FEMA flood maps. To reflect the actual expected damages, we follow the latter method and calculate EAD as the area under the EPL curve that represents damages against exceedance probability. EAD is defined as

$$EAD = \int_{P_{min}}^{P_{max}} D(p) \, dp , \qquad (6)$$

where $p$ is exceedance probability derived from GEV distribution. $D(p)$ is the damage caused by a flood with an exceedance probability of $p$. We calculate the damages using the depth-damage function.

Under the ignoring-uncertainty assumption, we derive $D$ from the HAZUS depth-damage function and the house lifetime is 30 years. Under this assumption, $p$ is from a GEV model, parameters of which are the maximum *a posteriori* likelihood estimations (the mode of the posterior distribution). Discount factor is

$$F_t = exp(-\sum_{s=0}^{t} r), \qquad (7)$$

with an $r$ value of 4 % per year.

Under the considering-uncertainty assumption, $O_{2h}$ becomes an ensemble and the mean of that ensemble is the expected total cost under uncertainty. Under uncertainty $O_{2h}$ becomes

$$O_{2h}^{unc} = E[O_{2h}^i] = E[C_h + LED_h^i], \qquad (8)$$

where

$$LED_h^i = \sum_{t=0}^{n} EAD_h^i * F_t^i. \qquad (9)$$

In these equations, $i$ indicates an index in the state space. Each state vector in the state space is called a State of the World (SOW). We create a state-space by random sampling (Supplementary Note 5). Samples are drawn from sources identified in section 4.2. In cases where the type of uncertainty is deep, we randomly switch samples from different models.

The elevations that minimize the total discounted costs with and without uncertainty are

$$h_{opt} = Arg\ Min_{h \in [0,14]}(O_{2h}), \qquad (10)$$

and

$$h_{opt}^{unc} = Arg\ Min_{h \in [0,14]}(O_{2h}^{unc}), \tag{11}$$

respectively.

In our cost-benefit analysis (CBA), the cost is the upfront cost ($C_h$) of elevating a house by $h$ feet. The benefits ($B_h$) are the net present value of the savings after elevating the house by $h$ feet. The benefit-to-cost ratio is $O_{3h} = \frac{B_h}{C_h}$ where $B_h = LED_h - LED_0$.

When uncertainty is ignored, we calculate LED using Eq. 5 with values discussed in the previous section. When uncertainty is considered, $O_{3h}$ becomes an ensemble. We use the mean of this ensemble as the expected benefit-to-cost ratio under uncertainty.

We define reliability as the probability of no flooding during the house lifetime. For a building that is elevated by $h$ feet, reliability is

$$O_{4h} = \prod_{t=1}^{n} \Pr Pr\ (X \leq h)\ = (CDF_h)^t, \tag{12}$$

where $n$ is the house lifespan and CDF denotes the probability that the annual maximum water level does not exceed the house's lowest level. Under uncertainty, reliability is the expected value of the ensemble of reliabilities for all SOWs.

Robustness is often measured using the concepts of satisficing and regret. Satisficing-based measures focus on outcomes that are within acceptable ranges defined for each objective. Regret-based criteria, on the other hand, focus on the deviations in performance caused by incorrect assumptions/decisions[28,29]. In this study, we assess the robustness of heightening strategies using a satisficing-based criterion called the domain measure[28]. This satisficing index measures the fraction of SOWs in which one or more objectives fall within the acceptable range. The acceptable ranges in our analysis are $[1,\infty)$ for the benefit-to-cost ratio, $[0,0.75]$ for the ratio of the total cost to house value, and $[0.5,1]$ for reliability.

### 4.4. Sensitivity analysis

We use global sensitivity analysis (GSA) to quantify the relative importance of uncertainty sources in determining expected damages[66]. Unlike the one-at-a-time (OAT)[67] sensitivity analysis approach that varies each factor separately, GSA allows variation of all the factors at the same time. This allows for understanding the effects of interactions between factors[41]. If $y=f(x_1,x_2,...,x_j,...,x_k)$, the relative importance of an individual factor ($x_j$) (also known as first-order sensitivity index) is $S_j = \frac{V(E(y|x_j))}{V(y)}$, which is the variance of the expected value of y conditioned on $x_j$ divided by the unconditional variance[66]. Sobol' sensitivity analysis identifies a subset of factors that accounts for most of the variance in output[68]. The total variance of the output is decomposed into elements that come from individual parameters and their interactions. Sobol''s first-order index indicates the effects of a single parameter on the model output. The total-order effect is the combination of the first-order effect and all the interactions with other parameters. Since Sobol''s method becomes computationally expensive in high parameter spaces, Saltelli's method, which uses fewer simulations, is often used for high-order indices[66]. Saltelli proposes two theorems[66]. The first theorem calculates the full set of first- and total-order indices at the computational cost of $n(k+2)$. The second theorem calculates first-, second-, and total-order indices at the cost of $n(2k+2)$, where *n* is the number of Monte Carlo samples and *k* is the number of parameters. In this study, we use Saltelli's second theorem to quantify the first-, second-, and total-order indices. We use the R package "sensitivity"[69].

**Data availability**

USGS water level and streamflow data can be accessed at [https://waterdata.usgs.gov/nwis/uv?site_no=01554000]. USGS rating curve can be accessed at [https://waterwatch.usgs.gov/?m=mkrc&sno=07050500]. Discount rate time series and all data used in this paper are available, under the GNU General Public License (version 3 or later), at [https://github.com/scrim-network/Zarekarizi-flood-home-elavate.git]

**Code availability**

All code used in this paper are available, under the GNU General Public License (version 3 or later) at [https://github.com/scrim-network/Zarekarizi-flood-home-elavate.git]

The code is written in R and is comprised of the following main steps: (1) converts streamflow USGS observations into water level and extracts annual maximums, (2) fits the statistical distribution to the annual maximums and estimates return levels, (3) analyzes all stakeholder objectives for a sample house with and without uncertainty quantification, (4) analyzes trade-offs between objectives, (5) repeats step 4 for a large sample of houses, and (6) conducts sensitivity analysis for all uncertainty and exposure scenarios.

**Acknowledgments**

This work was co-supported by the Penn State Initiative for Resilient Communities (PSIRC) by a Strategic Plan seed grant from the Penn State Office of the Provost, with co-support from the Center for Climate Risk Management (CLIMA), the Rock Ethics Institute,

Penn State Law, the Hamer Center for Community Design, and by the National Oceanic and Atmospheric Administration, Climate Program Office, under grant NA16OAR4310179. We thank Billy Pizer for sharing discount rate data. All errors and opinions are from the authors and do not reflect the funding agencies. We are grateful to James Doss-Gollin, Joel Roop-Eckart, David R. Johnson, Irene Schaperdoth, Tony Wong, Ben Lee, Murali Haran, Matthew D. Lisk, and Skip Wishbone for inputs.

**Disclaimer and License**

**Author contribution**

M.Z. and K.K. designed the study. M.Z. led the calculations, in collaboration with V.S. V.S. led the development of the discounting models and implemented them. M.Z. wrote the initial draft of the paper. All authors revised and edited the paper.

**Competing interests**

The authors declare no competing interests.

**Materials & Correspondence**

Correspondence and requests for materials should be addressed to M.Z.

**Supplementary**

**Neglecting Uncertainties Biases House-Elevation Decisions to Manage Riverine Flood Risks**

**Zarekarizi et al.**

**Supplementary Note 1:** We demonstrate the approach for a rural location in Pennsylvania (PA). Between 1959 and 2005, PA ranked 2nd, 10th, and 14th in the U.S. in the frequency of flash flood-related fatalities, injuries, and casualties, respectively[1]. In the same period, two from the ten deadliest events in the U.S. (excluding hurricane Katrina) have happened in PA, resulting in over 50 fatalities[1]. Within 1975 to 2019, FEMA paid $953 million to NFIP participants in Pennsylvania for property damages[2]. In response to these floods, some PA house owners have elevated their houses. Even though elevated to FEMA's standards, these houses still had over 12 million U.S. dollars in flood damages. Specifically, we choose Selinsgrove, a town by the Susquehanna River Basin, flowing into the Chesapeake Bay where frequent and severe floods are a major concern.

**Supplementary Note 2:** Given the localized nature of the house elevation decision, our analysis focuses on a specific case study but our approach is expandable and generalizable. What changes across locations are decisions about functional forms, considered values, or model parameters. The following notes might be necessary for case studies outside the U.S.

We use depth-damage functions that are originally provided by the U.S. Federal Emergency Management Agency (FEMA). For case studies outside the U.S. these functions need to be replaced.

We use three models to account for the uncertainty of the discount rate. These models are valid for case studies outside the U.S.; however, the models need to be calibrated based on historical interest rates for that country.

For house lifetime uncertainty, we use a PDF recommended based on the U.S. Census Bureau under the 2009 American Housing Survey microdata. For case studies outside the U.S., this PDF needs to be adjusted accordingly.

Depending on the location of the structure, it could be vulnerable to coastal, riverine, or compound flooding. In such cases, the flood hazard model used in this study needs to be calibrated based on historical flood observations specific to that location.

**Supplementary Note 3:** GEV distribution is often used for estimating water level distribution. For examples, see refs.[3-7].

**Supplementary Note 4:** Standard deviations of priors for location, scale, and shape parameters are 31.62, 10, and 1, respectively.

**Supplementary Note 5:** We use the Latin Hypercube Sampling method. For more information, see ref [8].

**Supplementary Table 1:** Uncertainties consider in this study, their types, and our approach in quantifying them. For deeply uncertain sources where we use multiple models, the more likely model is indicated in bold font. For details of each model, see the Methods section

| Uncertainty source | Uncertainty type | Uncertainty quantification method |
|---|---|---|
| Flooding frequency | shallow (one PDF) | We sample from GEV distribution |
| Depth-damage function | deep (two PDFs) | We use two distinct models (**HAZUS** and Huizinga et al., 2017) with 30% uniform error added to each. We consider HAZUS as the more likely scenario |
| Discount rate | deep (three PDFs) | We use three distinct models (a random walk model, a mean-reverting model, and a **mean-reverting model with background trend**). We consider the model with a background trend as the most likely model. |
| House lifetime | shallow (one PDF) | We sample from Weibull distribution |

**Supplementary Table 2:** Characteristics of the hypothetical pool of houses for the exposure study. We sample from independent uniform distributions bounded by the ranges below. We create a pool of 1,000 hypothetical buildings using Latin Hypercube Sampling

| Variable | Minimum | Maximum |
|---|---|---|
| House value ($) | 10,000 | 1,000,000 |
| House size (ft²) | 100 | 5000 |
| lowest level elevation with respect to BFE (ft) | -10 | 0 |

**Supplementary Table 3:** Estimated discount rate model parameter values. The standard deviations are provided in parentheses

| Parameter | Random Walk | Mean-Reverting | Background Trend |
|---|---|---|---|
| Mean | | 3.405 | |
| log-Mean standard error | | 0.3457 | |
| Intercept | | | 1.9289 (0.1728) |
| Trend | | | -0.0058 (0.0014) |
| AR1 | 1.7429 (0.0648) | 1.7371 (0.0649) | 1.6965 (0.0655) |
| AR2 | -1.0455 (0.1160) | -1.0270 (0.1175) | -0.9755 (0.1181) |
| AR3 | 0.3010 (0.0674) | 0.2806 (0.0710) | 0.2388 (0.0738) |
| $\sigma^2$ | 0.0034 | 0.0034 | 0.0033 |

**Supplementary Tables 4:** AIC and BIC of discounting models. The model with the lowest AIC and BIC is in bold

| Discount rate model | AIC | BIC |
|---|---|---|
| Random Walk | -617 | -603 |
| Mean-Reverting | -617 | -600 |
| Background Trend | **-624** | **-604** |

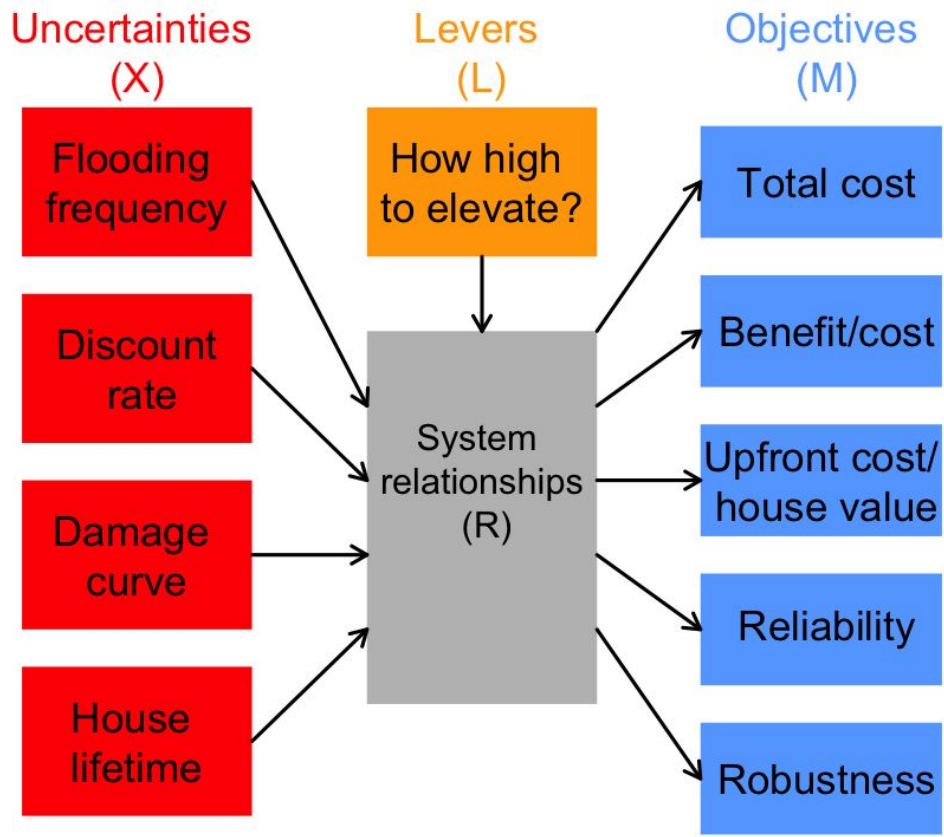

**Supplementary Figure 1:** An XLRM diagram that shows the decision framework. The orange element is the lever (L) (i.e. how high to elevate a house). Red components represent exogenous uncertain factors (X) that impact the decision and are out of control of the decision-maker. Objectives or metrics (M) represent how success is measured. System relationships (R) shows how levers and uncertainties translate into objectives

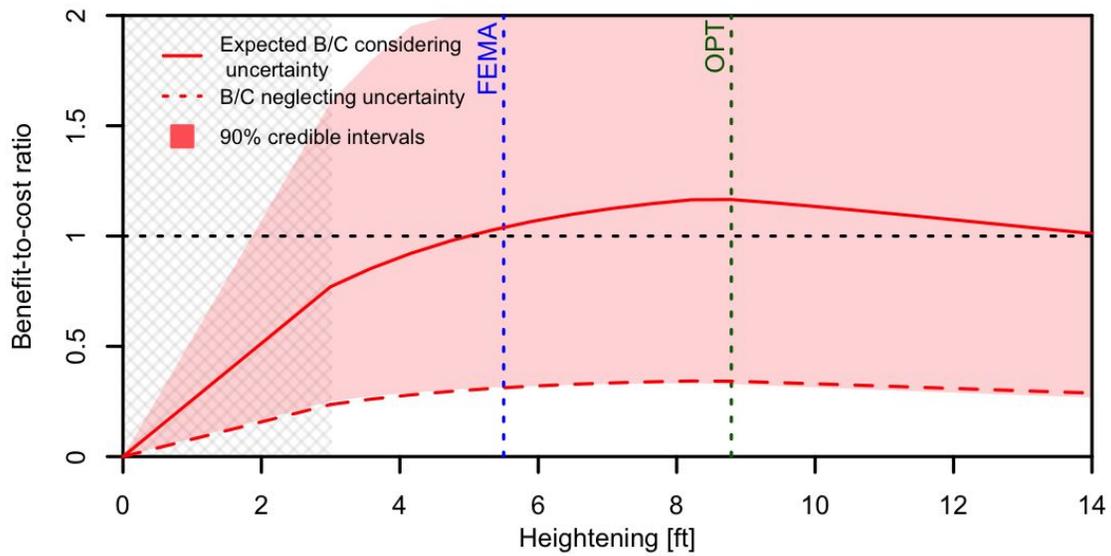

**Supplementary Figure 2:** The benefit-to-cost ratio under assumptions of ignoring-uncertainty and considering-uncertainty for the typical house studied in this paper. The blue vertical line indicates the FEMA-recommended heightening strategy. The green vertical line indicates the strategy recommended by the considering-uncertainty assumption. The hatched gray area on the left refers to elevating the house by less than three feet which we ignore in this study

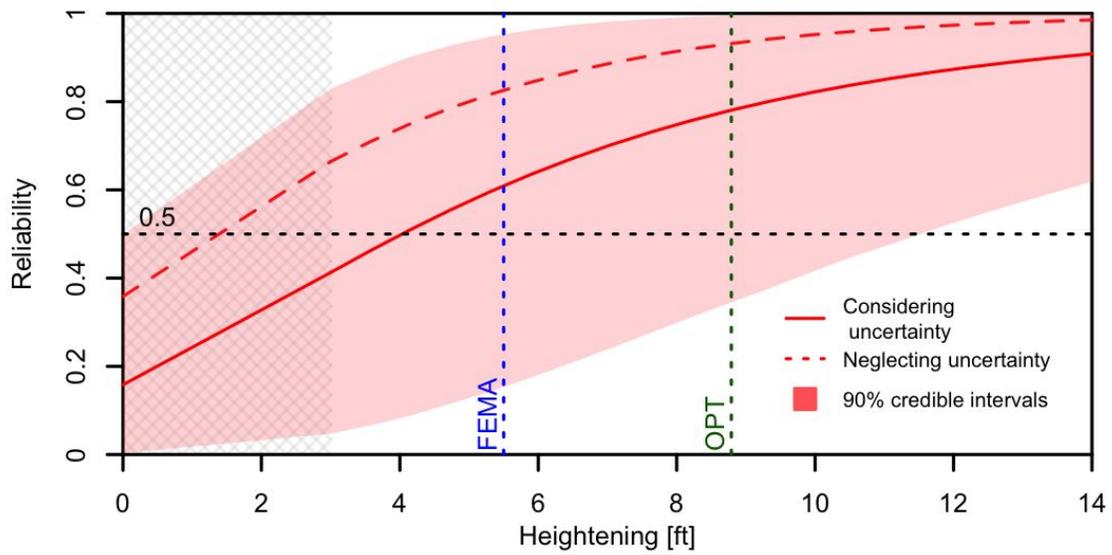

**Supplementary Figure 3:** Reliability under assumptions of ignoring-uncertainty and considering-uncertainty for the typical houses studied in this paper. The vertical line indicates the FEMA-recommended heightening strategy. The green vertical line indicates the strategy recommended by the considering-uncertainty assumption. The hatched gray area on the left refers to elevating the house by less than three feet which we ignore

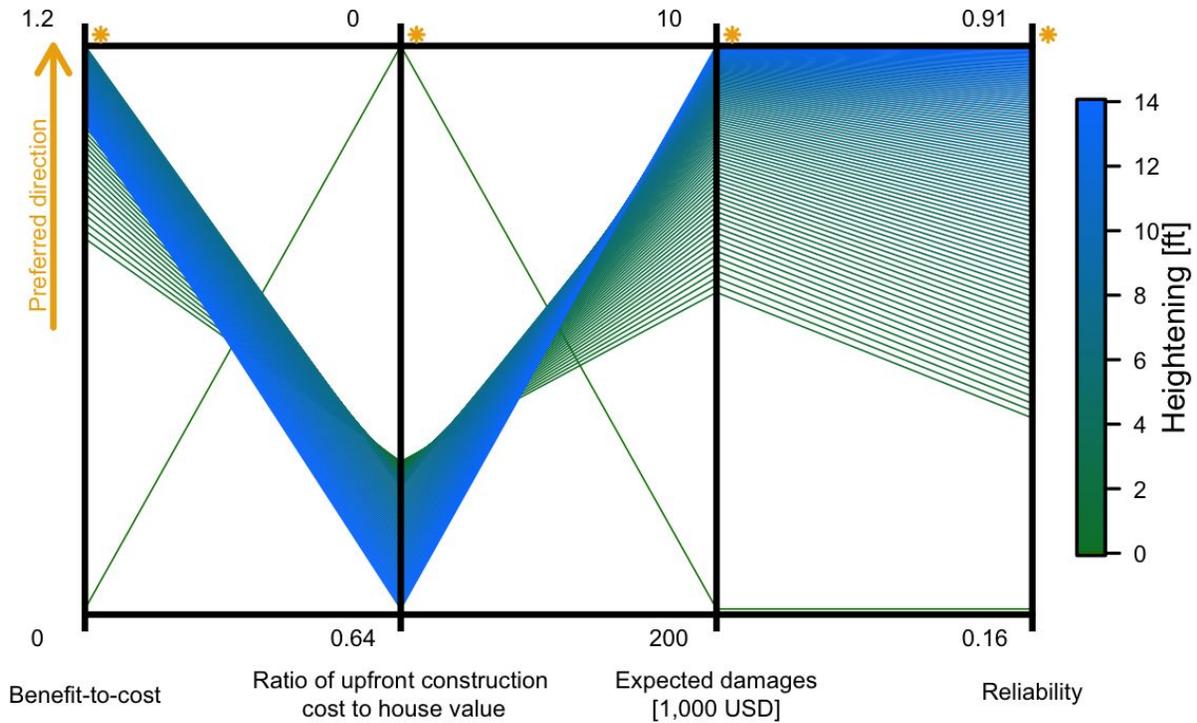

**Supplementary Figure 4:** Trade-offs between different decision-makers' preferences. Each line indicates a heightening policy. The left-out line indicates the not-elevating policy (a policy recommended by the ignoring-uncertainty assumption). The infeasible ideal policy yields a horizontal line on the top of the axes. Green lines represent lower lifting policies and blue lines indicate higher lifting policies. Policies with high (low) reliability are associated with low (high) expected damages, high (low) upfront costs, and high (low) benefit-to-cost ratio

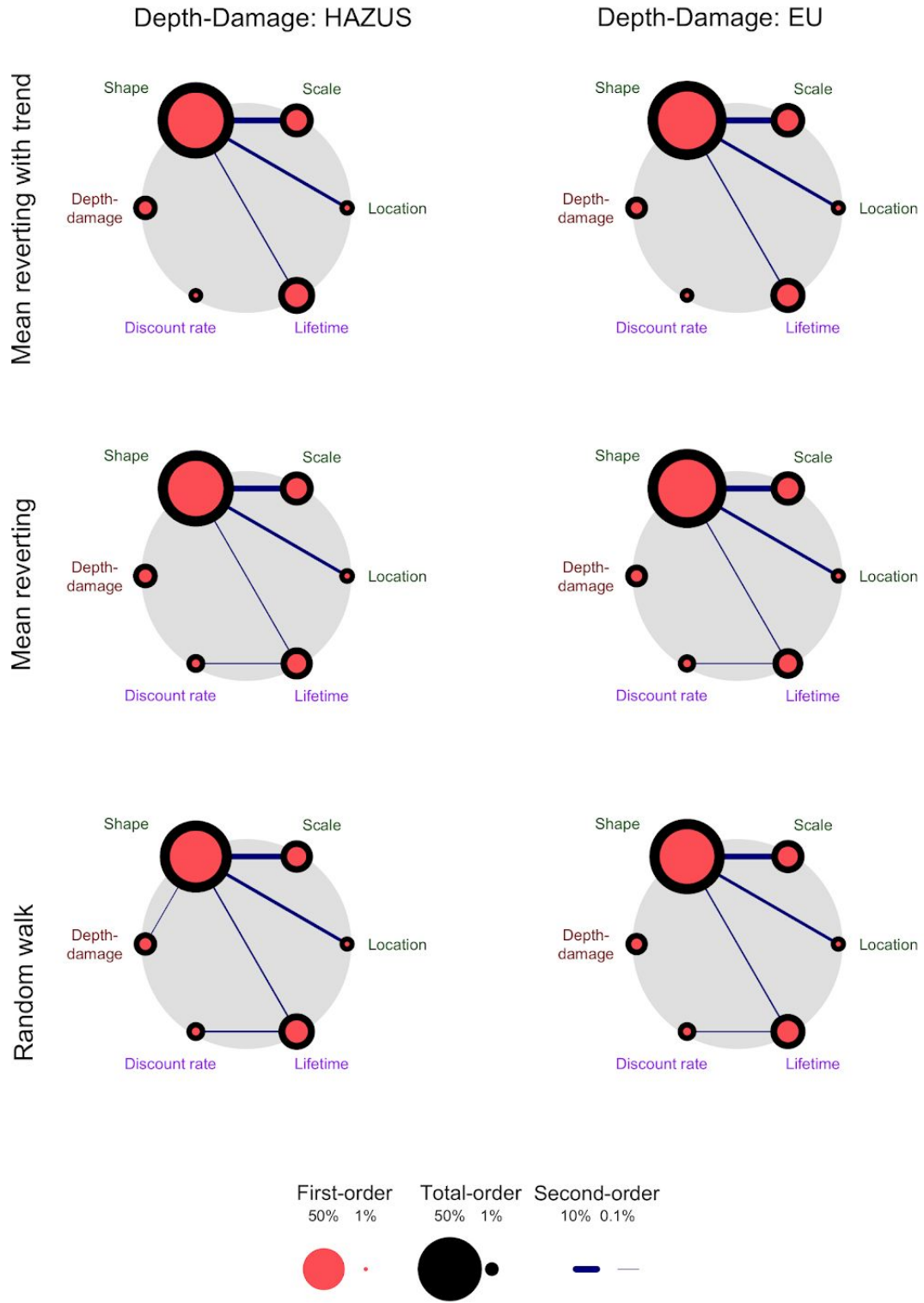

**Supplementary Figure 5:** Similar to Figure 5 but for different scenarios. Scenarios are defined based on combinations of discount rate and depth-damage model options

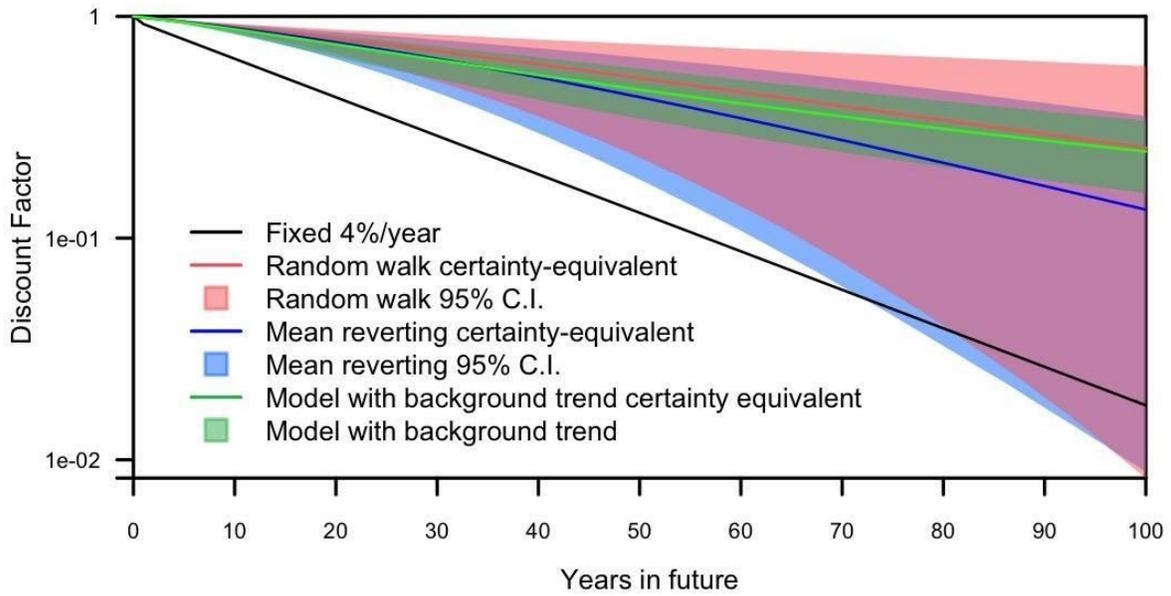

**Supplementary Figure 6:** 100-year-period discount factors of three stochastic models as compared with the discount factor of a constant positive discount rate. Shaded bounds indicate the uncertainties in the stochastic models

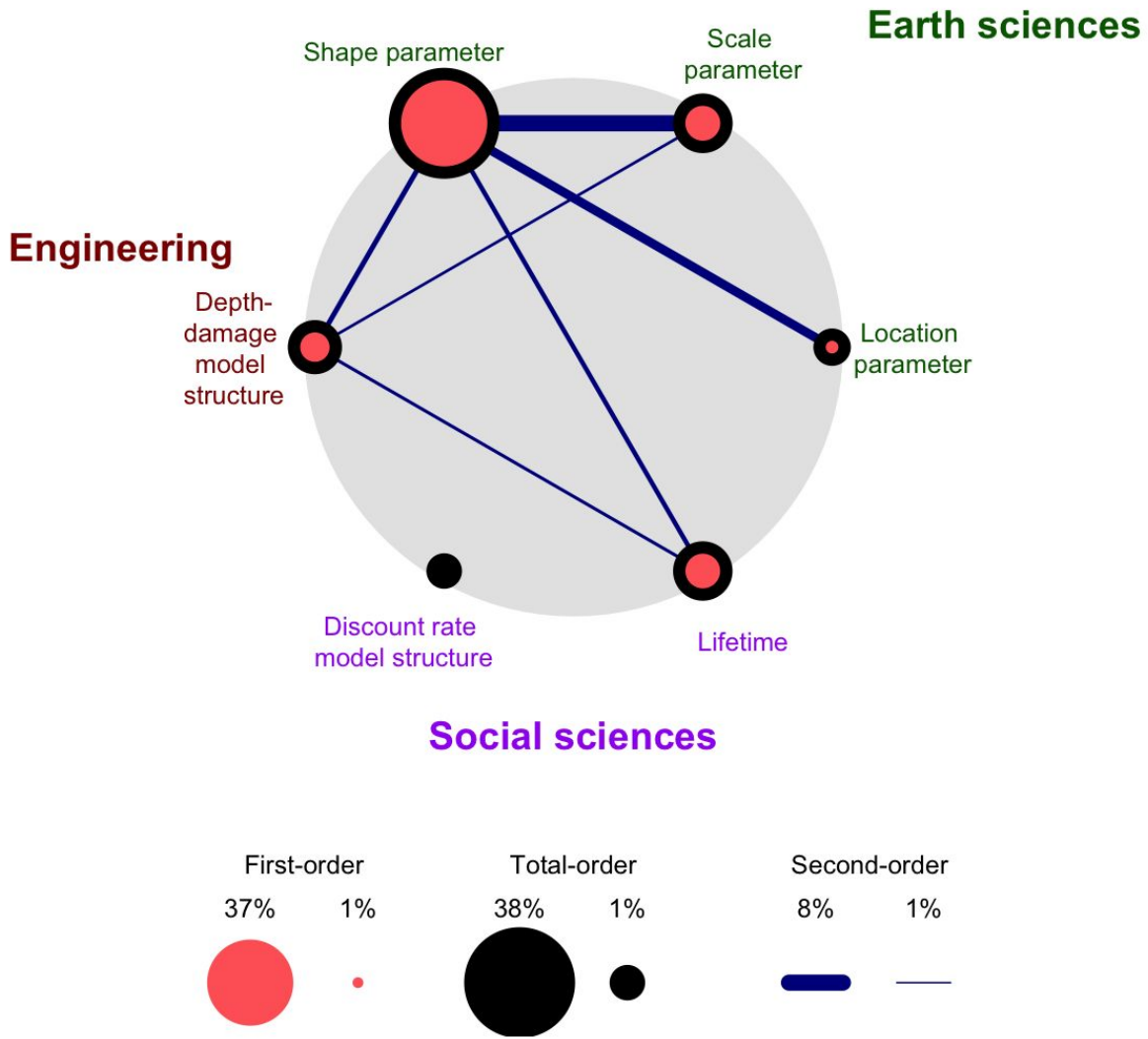

**Supplementary Figure 7:** Same as Figure 5 but for deep uncertainties. Here, the discount rate node indicates the model structure uncertainties. Samples for this node are drawn uniformly from the vector of (1,2,3). Each element represents a model choice. For depth-damage function, samples are drawn uniformly from two model choices as discussed in the methods

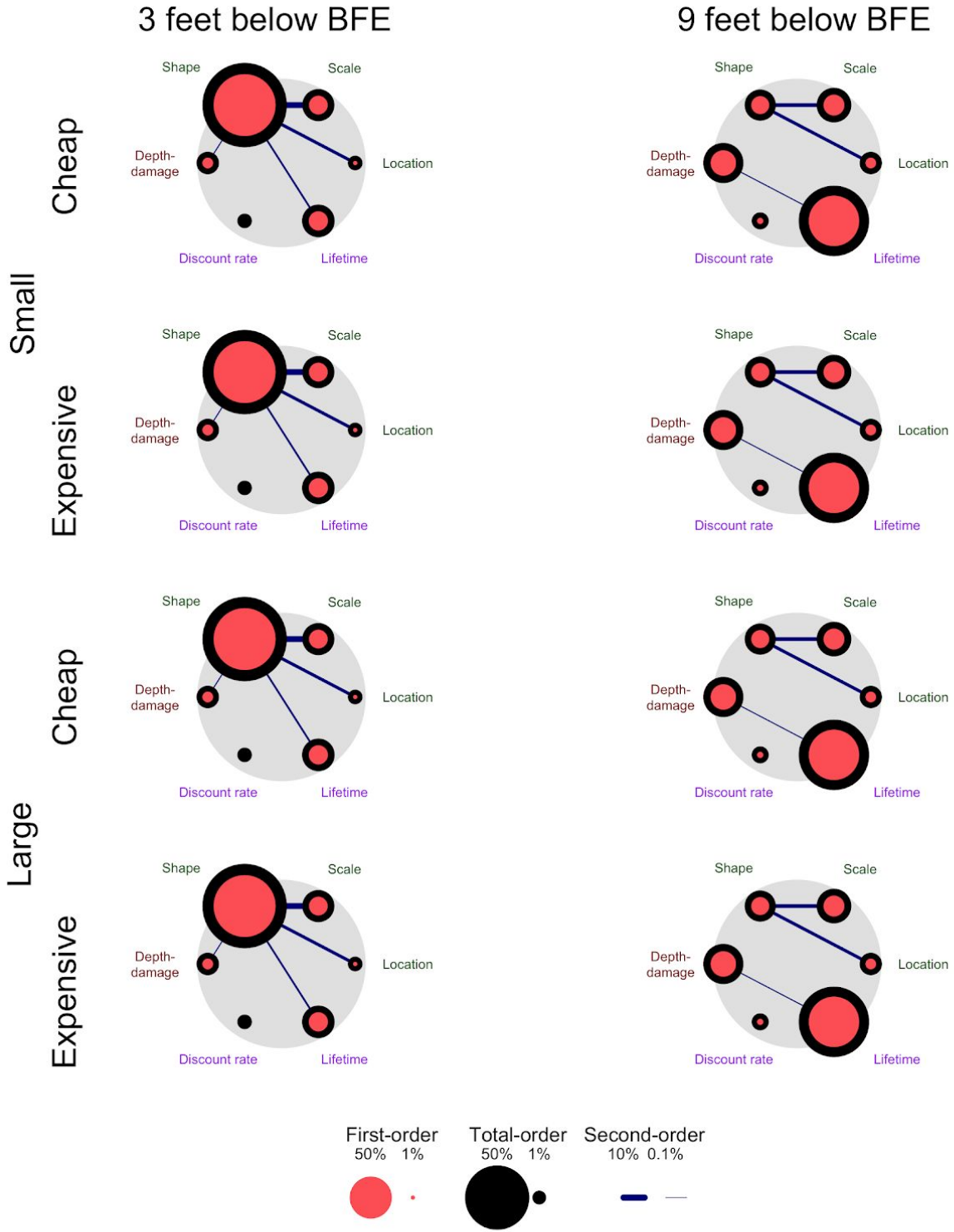

**Supplementary Figure 8:** same as Figure 5 but for different house exposure factors such as size, value, and the lowest level elevation. Small: 500 ft$^2$ large: 3,000 ft$^2$ cheap:$100,000 expensive:$600,000

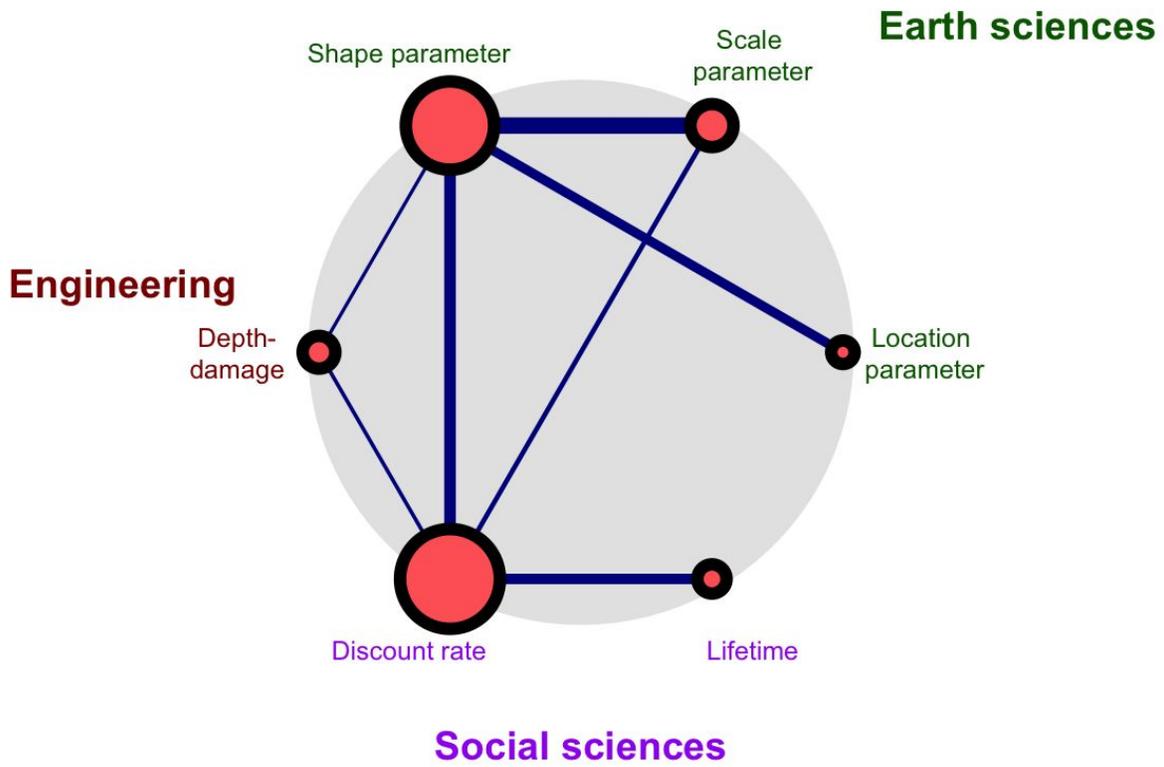

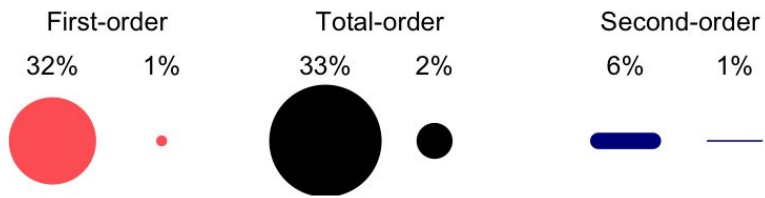

**Supplementary Figure 9:** Same as Figure 5 but with a different sampling approach for the discount rate. Here, we draw samples randomly from the [1%,10%] range

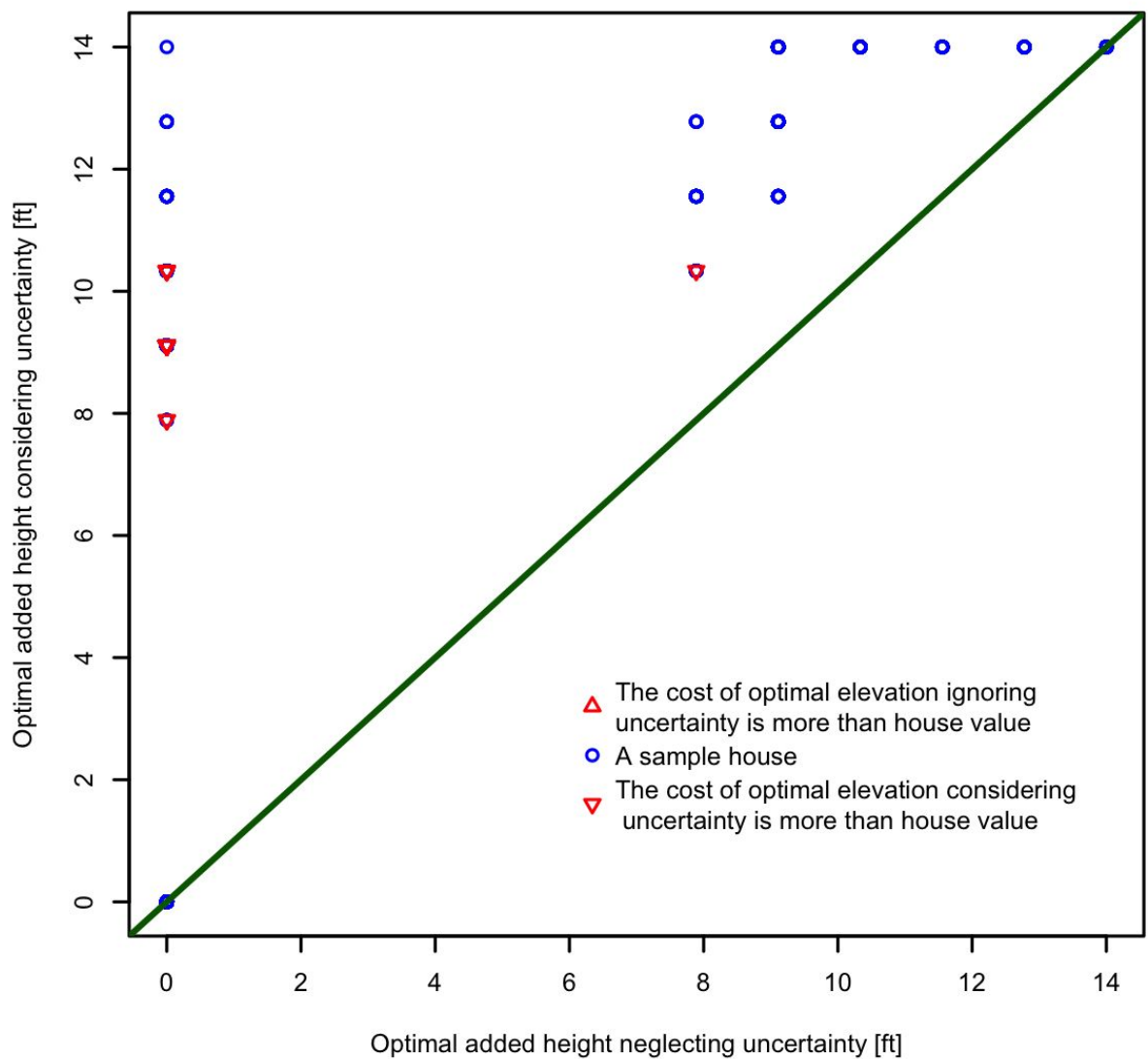

**Supplementary Figure 10:** Comparison of economically-optimal elevations under two assumptions of ignoring-uncertainty and considering-uncertainty. Each point represents a house. Houses in which one or both of the optimal elevations are more than house value are indicated by red

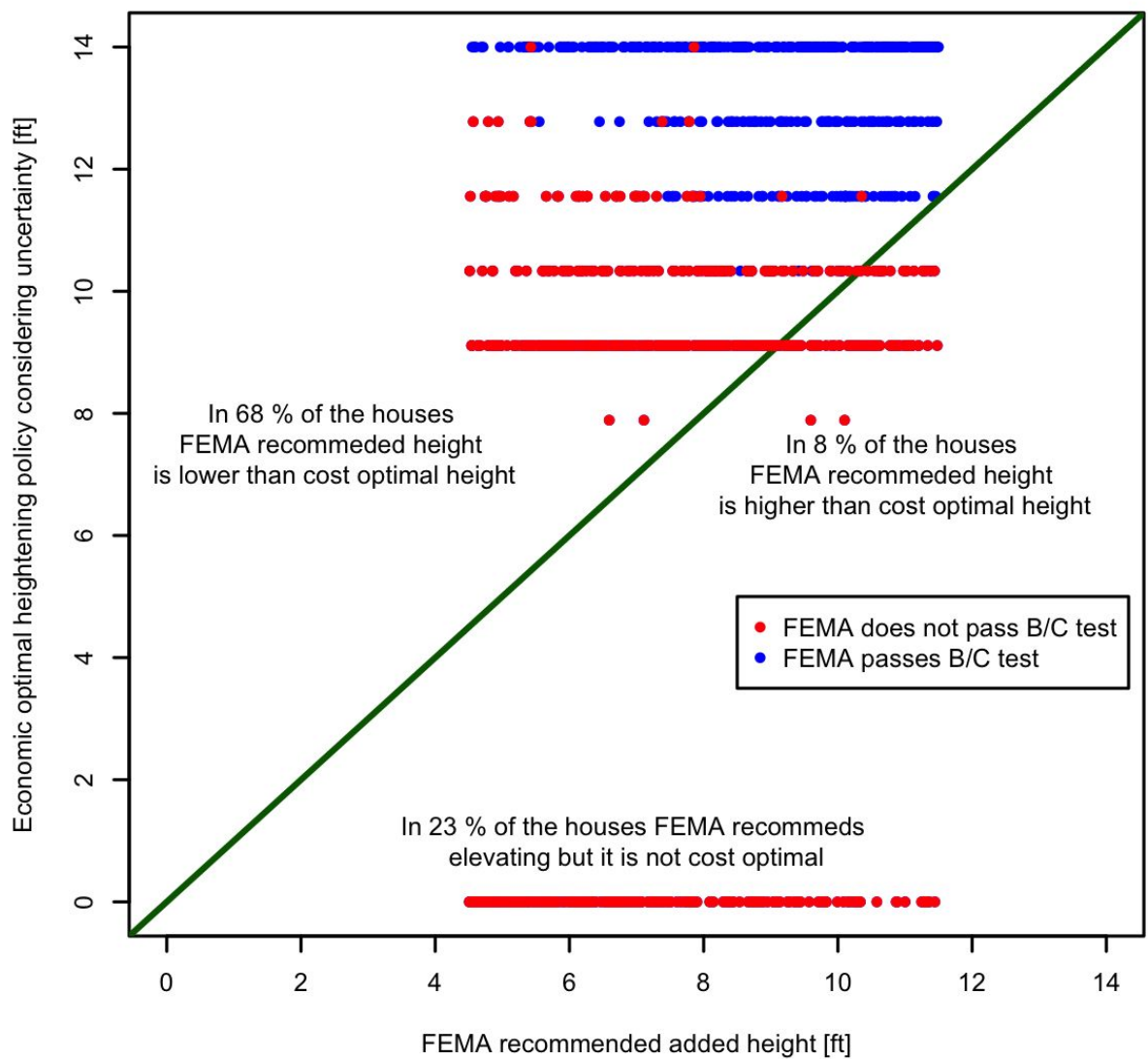

**Supplementary Figure 11:** The economic optimal elevation versus FEMA's recommendation. Each dot represents a house (a total of 1,000 houses). Red dots indicate that FEMA's recommended policy does not pass the cost-benefit test (i.e. the benefit is less than the cost). The diagonal green line is the 1:1 line

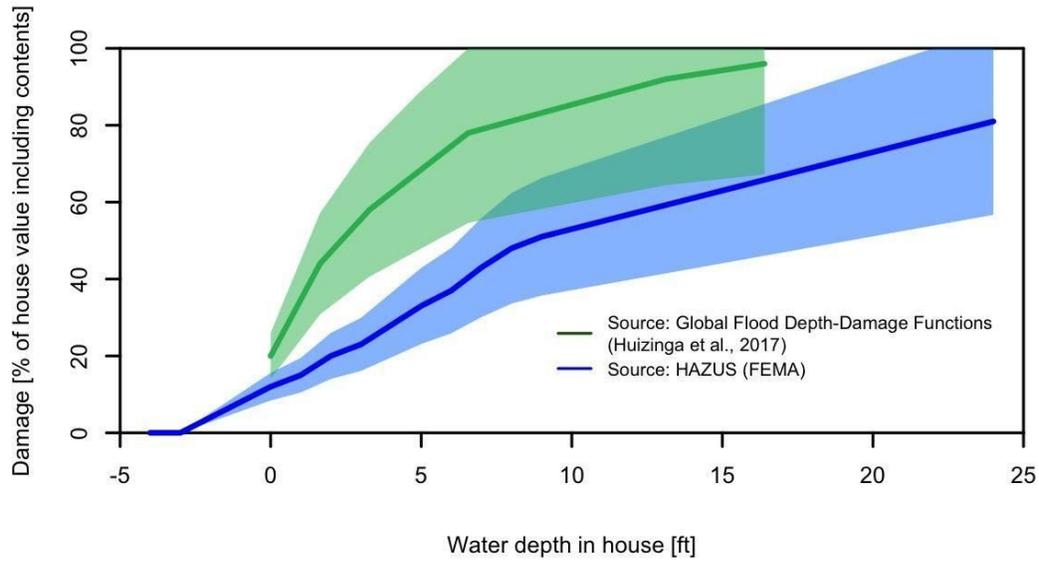

**Supplementary Figure 12:** Two depth-damage functions used in this study. The damage model in blue is obtained from FEMA HAZUS and the damage curve shown in green is obtained from combining multiple functions from HAZUS[8]. Shallow uncertainty in each function is represented by 30% uniform bounds (shown in light blue and green)

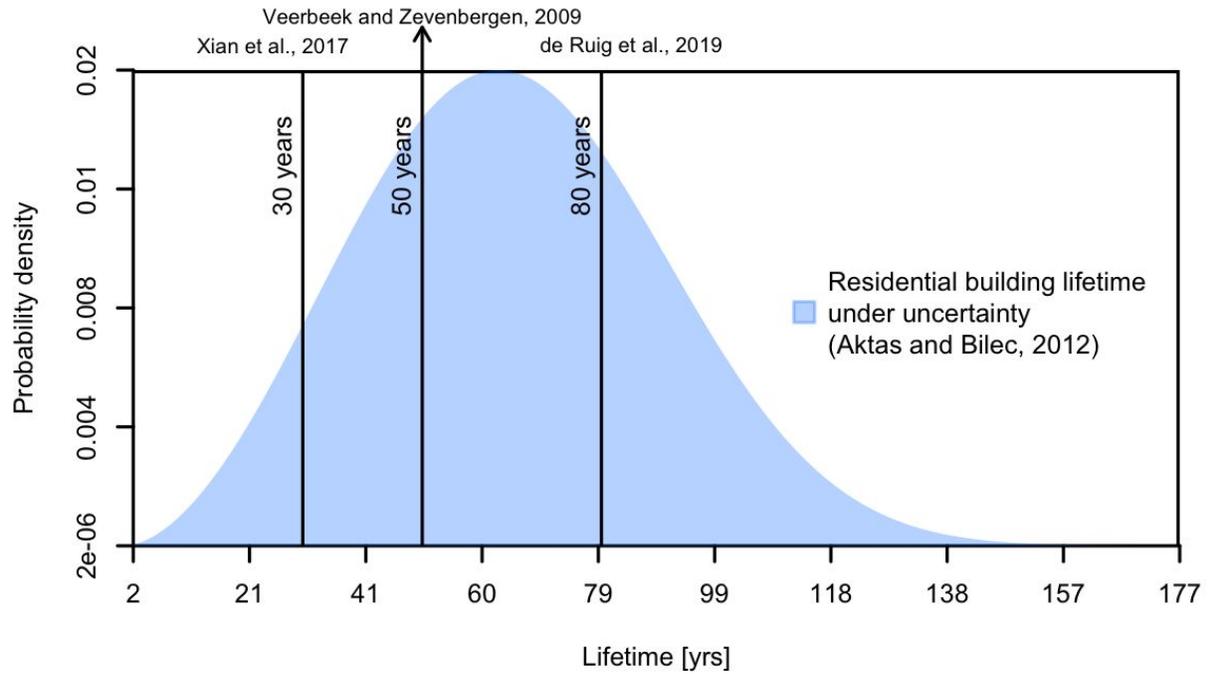

**Supplementary Figure 13:** The uncertainty in house lifetime considered in this study (the shaded blue distribution) and some deterministic values commonly used in the literature[9-11] (vertical black lines). The distribution is Weibull with shape and scale parameters of 2.8 and 73.5, respectively

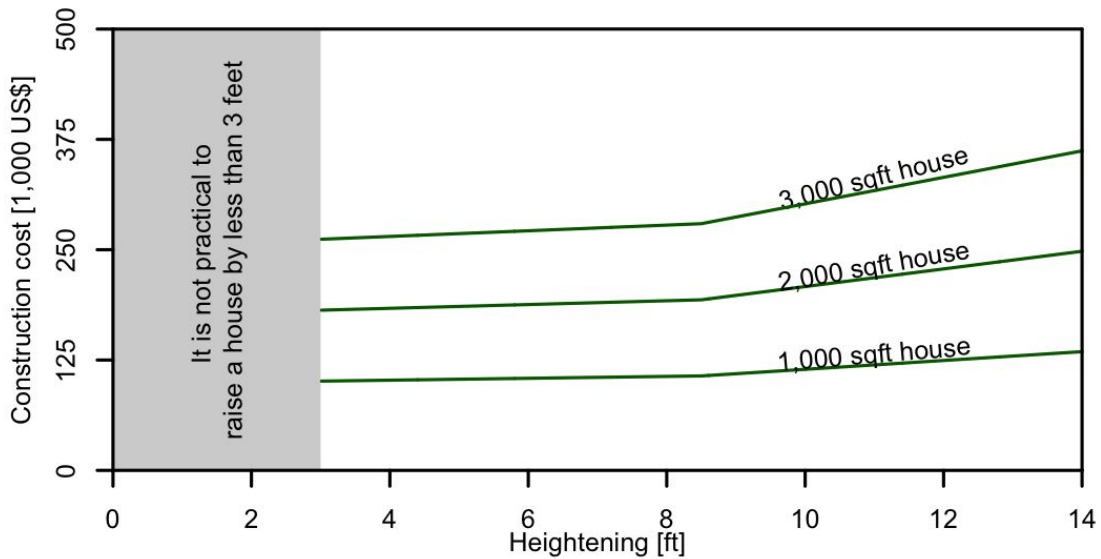

**Supplementary Figure 14:** Construction cost for three sample houses with sizes of 1,000, 2,000, and 3,000 ft$^2$. The gray area indicates an elevation of fewer than three feet which we assume to be impractical. These cost estimates are adopted from the CLARA model. Units are in 2017 US$ value

**Supplementary References:**